%% file: Tensor_dS3.tex
\documentclass[a4paper,onecolumn,preprintnumbers,nofootinbib]{revtex4}

\usepackage[top=2.8cm, bottom=2.8cm, left=2.4cm, right=2.4cm]{geometry}
\usepackage[utf8]{inputenc}  
\usepackage[T1]{fontenc}     
\usepackage{amsmath,amssymb,lmodern} 
\usepackage{graphicx}
\usepackage{hyperref}
\usepackage{color}

\input{shortcuts}

\begin{document}
\preprint{PI/UAN-2019-648FT}

\title{Correlation functions of sourced gravitational waves in inflationary scalar vector models. A symmetry based approach}

\author{Juan P. Beltr\'an Almeida$^1$}
\email{juanpbeltran@uan.edu.co}
\author{Josu\'e Motoa-Manzano$^2$}
\email{josue.motoa@correounivalle.edu.co}
\author{C\'esar A. Valenzuela-Toledo$^2$}
\email{cesar.valenzuela@correounivalle.edu.co}
\affiliation{$^1$Departamento de F\'isica, Facultad de Ciencias, Universidad Antonio Nari\~no, \\ Cra 3 Este \# 47A-15, Bogot\'a DC, Colombia}
\affiliation{$^2$Departamento de F\'isica, Universidad del Valle, \\
Cll. 13 \# 100 - 0 A.A. 24360, Cali, Colombia}
\date{\today}

\begin{abstract}

We use conformal symmetry to constrain the shape of inflationary correlators in the presence of long-lived vector field perturbations. Applying conformal Ward identities, we derive general expressions, up to amplitudes and normalization factors, for the two and three-point correlators in the presence of vector fields mediated by the interaction $f(\phi)\left(F_{\mu \nu}F^{\mu \nu}+\alpha\tilde{F}_{\mu \nu}F^{\mu \nu}\right)$, where $f(\phi)$ is a suitable coupling function between the scalar and the vector field. This interaction allows for isotropy and parity symmetry breaking and is consistent with super horizon conformal symmetry. As an application of the conformal field theory techniques followed here, we evaluate the mixed tensor-scalar  $\langle \gamma \zeta  \rangle$ and tensor-scalar-scalar $\langle   \gamma \zeta \zeta \rangle$ correlators which are interesting to look for parity violating effects related with chiral gravitational waves. Finally, we derive consistency relations for the three-point correlators obtained. 
 
\end{abstract}

\maketitle

\section{Introduction}
The relation between the de Sitter space and the conformal field theory have been widely used in cosmology. 
In particular, a powerful application of this relationship is to use the isomorphism between the de Sitter group in four dimensions and the conformal group in three dimensions to fix the form of the primordial correlation functions \cite{Antoniadis:1996dj, Strominger:2001pn, Maldacena:2011nz, Antoniadis:2011ib, Creminelli:2011mw, Creminelli:2012ed, Kehagias:2012pd, Kehagias:2012td, Mata:2012bx, Coriano:2013jba,   Bzowski:2013sza, Bonora:2014qla, Bonora:2015nqa, Isono:2016yyj, Kehagias:2017rpe, Bzowski:2018fql}\footnote{Another approach to the use of symmetries to determining the form of cosmological correlators can be found in \cite{Marcori:2016oyn}.}. The key idea of this method comes from the fact that, during the inflationary expansion, deep in the super horizon regime, the  fields behave (approximately) like conformal fields.   
This fact allows us to use  conformal field theory (CFT) methods, such as Ward identities associated with the conformal symmetries of the theory, as a powerful tool to study general properties of the inflationary correlation functions.  

On the other hand, CMB observations do not totally rule out the existence of small deviations from the standard predictions of the usual $\Lambda$CDM cosmological model, in particular, there is some (small) room for cosmological models predicting non-vanishing levels of  
non-gaussianity, parity violation signals, statistical anisotropy patterns, among others \cite{Akrami:2018odb, Ade:2015hxq, Ade:2015ava,Perivolaropoulos:2014lua}.  Some of these deviations, commonly referred as anomalies, could be explained with the presence of vector fields (also with the presence of higher spin fields \cite{Arkani-Hamed:2015bza,Kehagias:2017cym,Bartolo:2017sbu,Franciolini:2017ktv,Baumann:2017jvh,Franciolini:2018eno,Bordin:2018pca,Anninos:2019nib} or $p-$forms \cite{Koivisto:2009sd, Koivisto:2009fb, Mulryne:2012ax, Ohashi:2013qba, Ohashi:2013mka, Kumar:2016tdn, Obata:2018ilf, Almeida:2018fwe, Almeida:2019xzt}) during the inflationary expansion. A vector field with a non zero vacuum expectation value (VEV)  constitutes a natural source of anisotropies and parity breaking signatures because the VEV introduces a privileged direction in the CMB map.  
Among the various cosmological models that include vector fields  
(see \cite{Dimastrogiovanni:2010sm, Soda:2012zm, Maleknejad:2012fw} for reviews on the topic), we are interested here in a particular class of models in which a single $U(1)$ invariant vector field is coupled to a scalar field through non standard kinetic couplings of the form 
 \begin{equation}\label{lphia}
{\cal L}_{A}(\phi, A) = f_1(\phi) F_{\mu\nu}F^{\mu\nu} +  f_2(\phi) F_{\mu\nu}\tilde{F}^{\mu\nu}, 
\end{equation}  
where $f_{1,2}(\phi)$ are functions of the scalar field, $F_{\mu\nu} = \partial_{[\mu} A_{\nu]}$ is the field tensor of the vector field, and $\tilde{F}_{\mu \nu} = \frac{1}{ 2\sqrt{-g}} \varepsilon_{\mu \nu \alpha \beta} F^{\alpha \beta}$ is its Hodge dual. Within the context of inflation, many aspects of this class of models have been studied with great detail over the last years  
\cite{Ratra:1991bn, Yokoyama:2008xw, Watanabe:2009ct, Dimopoulos:2009vu, Dimopoulos:2009am, Anber:2009ua, Watanabe:2010bu, Sorbo:2011rz, Dimopoulos:2012av, Anber:2012du, Bartolo:2012sd, Barnaby:2012xt, Biagetti:2013qqa, Durrer:2013pga, Cook:2013xea, Shiraishi:2013vja, Shiraishi:2013kxa, Abolhasani:2013zya, Lyth:2013kah, Rodriguez:2013cj, Lyth:2013vha, Shiraishi:2013oqa, Chen:2014eua, Almeida:2014ava, Bartolo:2014hwa,Fleury:2014qfa, Caprini:2014mja, Bartolo:2015dga, Namba:2015gja, Shiraishi:2016yun, Fujita:2017lfu, Caprini:2017vnn, Almeida:2018pir}.  A specific case of this model results when the two kinetic coupling functions are proportional to each other $f_2(\phi) = \alpha f_1(\phi)$ \cite{Dimopoulos:2012av,Caprini:2014mja, Bartolo:2015dga, Valenzuela2016, Almeida:2017lrq, Caprini:2017vnn}.  Besides leading to interesting possibilities such as providing a mechanism for the generation of the seeds of inflationary magnetogenesis \cite{Caprini:2014mja,Caprini:2017vnn}, and the prediction of non-diluting statistical anisotropic and parity breaking patterns \cite{Dimopoulos:2012av,Bartolo:2015dga}, the vector perturbations in this particular case preserve conformal invariance at super horizon scales when the kinetic coupling function is a homogeneous function of the scale factor \cite{Valenzuela2016, Almeida:2017lrq}. The conformal invariance of the vector perturbations offers the possibility of studying this scalar-vector model using CFT techniques to elucidate the structure of inflationary correlation functions. CFT was applied to the study of the inflationary correlators in presence of vector perturbations in the model $f(\phi) F_{\mu\nu}F^{\mu\nu}$ in \cite{Biagetti:2013qqa} and, subsequently, this methodology was applied to the case of the model $f(\phi) (F_{\mu\nu}F^{\mu\nu} +  \alpha F_{\mu\nu}\tilde{F}^{\mu\nu})$ in \cite{Valenzuela2016, Almeida:2017lrq}.  These works consider general implications of the conformal symmetry for the correlators involving scalar and vector perturbations, but, they don't consider tensor perturbations in their analysis.   

Our main purpose in this paper is to extend the application of CFT to derive general expressions for the inflationary correlators in Fourier space involving tensor perturbations/gravitational waves $\gamma_{ij}$ sourced by the vector fields. To fulfill this objective, we apply the vast results available in the literature about CFT in momentum space. To our knowledge,  the most comprehensive study of the implications of conformal symmetry in momentum space in the literature can be found in Ref. \cite{Bzowski:2013sza}.  This reference  displays a complete list of results for the ``3-point functions of the stress-energy tensor, conserved currents and scalar operators'' obtained by solving the Ward identities related with the conformal symmetries, basically, by solving the conformal Ward identities of dilatations and special conformal transformations.  Given that Ref. \cite{Bzowski:2013sza} is devoted to the analysis of parity conserving systems\footnote{Parity breaking models in momentum space have been considered for instance in \cite{Bonora:2014qla,Bonora:2015nqa}.}, the main contribution that we aim to present here is to extrapolate the results obtained in \cite{Bzowski:2013sza} to inflationary scalar-vector models in the presence of a parity breaking term of the form $ f(\phi) F_{\mu\nu}\tilde{F}^{\mu\nu}$. The main difficulty that we found to this purpose, was to find a complete extension of the tensorial decomposition for correlators involving vectors and tensors obtained in such reference, in such a way that the contributions from the parity breaking sources were properly taken into account. Here, we perform a careful search of the tensors that account for the presence of  parity breaking sources, 
necessary to construct the 3-point correlators of gravitational waves and scalar perturbations sourced by the vector perturbations coming from the interaction $f(\phi) (F_{\mu\nu}F^{\mu\nu} +  \alpha F_{\mu\nu}\tilde{F}^{\mu\nu})$. Remarkably, what we found is that the parity breaking terms in the tensor decomposition satisfy  the same differential equations implied by the conformal Ward identities, making the results of \cite{Bzowski:2013sza} directly applicable to our case. We use this fact to write a general parametrization of the 2-point functions $\langle \zeta \zeta  \rangle$, $\langle \gamma \zeta  \rangle$  and $\langle \gamma \gamma  \rangle$ and the 3-point functions $\langle   \zeta \zeta \zeta \rangle$, $\langle   \gamma \zeta \zeta \rangle$. Then, we study the soft limits (the limit configuration in which one of the momenta is going to zero) of the resulting 3-point functions, which become useful for observational purposes.
 
This work is organized as follows: in section \ref{eom} we describe the scalar-vector model consistent with super horizon conformal invariance of the vector perturbations, and describe the asymptotic behavior of the fields in the super horizon region. Then, in section  \ref{spectrum}, we compute the form of the two point functions for the scalar-scalar, the scalar-tensor  and the tensor-tensor perturbations by including the perturbative effect of the vector fields. In this section we also introduce the necessary structures accounting for the parity breaking tensor decomposition of the correlators. We  calculate the vacuum and ``vector fields sourced''  scalar-scalar-scalar and tensor-scalar-scalar 3-point correlation functions in section \ref{bispectrum}. In section \ref{BSST} we discuss the soft limit of the bispectrum calculated in section \ref{bispectrum}. Finally, in section \ref{Conclusions} we write our final comments and conclusions. 

\section{Scalar-vector model}\label{eom}

In this work, we study a cosmological scenario, in which a single $U(1)$ invariant vector field  $A_{\mu}$ is coupled to a single scalar field $\phi$, not necessarily the inflaton, during the inflationary period. The relevance of vectors, or higher spin, fields during inflation are severely constrained by current observations and, on theoretical grounds, by the ``cosmic no-hair theorem'' \cite{Wald:1983ky} which, in a broad sense, states that, all the possible imprints left by such fields dilutes exponentially fast during inflation, making them unobservable and irrelevant.  Nevertheless, a way to bypass the conditions of the no-hair theorem, in the case of $U(1)$ vector fields, consists in the introduction of suitable field dependent couplings of the form $f_1(\phi) F_{\mu\nu} F^{\mu\nu} $ and $f_2(\phi) F_{\mu\nu}\tilde{F}^{\mu\nu}$, in such way that vector field perturbations survive the inflationary dilution rate and become long-lived at superhorizon scales. We will focus here in the class of inflationary scalar-vector models, in which the scalar-vector interactions are mediated by the couplings aforementioned.
The full action of the model can be written as 
\ba\label{action0}  
S &=& \int d^{4}x \sqrt{-g}\left[  \frac{M_{pl}^{2}}{2} R - \frac{1}{2} \partial_{\mu}\phi \partial^{\mu}\phi - V(\phi)-\frac{1}{4} f_1(\phi)F_{\mu\nu}F^{\mu\nu}  -\frac{1}{4} f_2(\phi) F_{\mu\nu}\tilde{F}^{\mu\nu} \right],
\ea
where $f_1(\phi), f_2(\phi)$ are suitable functions of the scalar field.  
As mentioned previously in the introduction, the models that falls into this category have been widely studied in the recent literature and many issues, such as the stability and causality of the model, the sourcing of statistical anisotropic and parity breaking patterns in the CMB map, the production  of chiral gravitational waves, inflationary magnetogenesis, among many others issues of current interest, have been discussed with great detail \cite{Ratra:1991bn, Yokoyama:2008xw, Watanabe:2009ct, Dimopoulos:2009vu, Dimopoulos:2009am, Anber:2009ua,Watanabe:2010bu, Sorbo:2011rz, Dimopoulos:2012av, Anber:2012du, Bartolo:2012sd, Barnaby:2012xt, Biagetti:2013qqa, Durrer:2013pga, Cook:2013xea, Shiraishi:2013vja, Shiraishi:2013kxa, Abolhasani:2013zya, Lyth:2013kah, Rodriguez:2013cj, Lyth:2013vha, Shiraishi:2013oqa, Chen:2014eua, Almeida:2014ava, Bartolo:2014hwa,Fleury:2014qfa, Caprini:2014mja, Bartolo:2015dga, Namba:2015gja, Shiraishi:2016yun, Fujita:2017lfu, Caprini:2017vnn, Almeida:2018pir}. Among the models included in the action \eqref{action0}, here, we restrict to the particular case, in which both coupling functions are proportional, so  $f_1(\phi)= f(\phi), \, f_2(\phi)=\alpha f(\phi)$, where $\alpha$ is a dimensionless constant \cite{Dimopoulos:2012av,Caprini:2014mja, Bartolo:2015dga, Valenzuela2016, Almeida:2017lrq,Caprini:2017vnn}. In this case, the interactions in the scalar-vector  sector are described by the interaction Lagrangian: 
\be \label{coup}
{\cal L}_{int} = -\frac{1}{4}f(\phi)(F_{\mu\nu}F^{\mu\nu} +  \alpha F_{\mu\nu}\tilde{F}^{\mu\nu}).
\ee
Our main interest for such particular case resides on the fact that, when the coupling function $f(\phi)$ is a homogeneous function of time, deep in the super horizon regime  $k/(aH) \rightarrow 0$, the generated vector perturbations  preserve conformal invariance \cite{Valenzuela2016,Almeida:2017lrq}.   
This fact becomes very useful because the inflationary coupled scalar-vector system can be analyzed by using the vast quantity of tools, techniques and results of CFT.  The main reason why we use conformal symmetry in this context is because the spacetime background, during the inflationary period, can be modelled very accurately with a four dimensional de Sitter space. The symmetries associated with this maximally symmetric spacetime, the de Sitter group, acts on the super horizon regime in the very same way as the conformal group in one dimension less. 

In the next subsection we review some basics of this model, in particular, we solve the equations of motion and describe its asymptotic super horizon limit. Afterwards, we discuss the way in which the vector field coupled through \eqref{coup} induces the perturbations in the curvature perturbation and in the gravitational waves. Further details of the mode solutions can be found in \cite{Dimopoulos:2012av,Caprini:2014mja, Bartolo:2015dga, Valenzuela2016, Almeida:2017lrq,Caprini:2017vnn}.  

\subsection{Asymptotic super horizon evolution of the modes}
%
Here we consider the super horizon limit  of the solutions to the equations of motion derived from  \eqref{action0}. 
In conformal planar coordinates, the line element becomes 
\be\label{tauflat}
ds^2 = \frac{1}{H^2 \tau^2} \left[ -d \tau^2 + \delta_{ij} dx^{i}dx^{j} \right], \qquad -\infty<\tau<0,
\ee
where $H$ is the Hubble parameter and $ \tau $ is the conformal time.  Let's consider first the case of a massive scalar field with mass $m$, propagating in de Sitter space. In the coordinates \eqref{tauflat}, the equation of motion for this field becomes 
\be\label{eqphi}
 {\phi}''-\frac{2}{\tau} {\phi}'-\nabla^2 \phi+\frac{m^2\phi}{H^2\tau^2}=0.
\ee 
The asymptotic expression for the growing mode (the dominant solution at late times) of this field, in the super horizon limit $k/(aH) = - k\tau \rightarrow 0$, can be written as (see {\it e.g.} section 6.2 of \cite{Riotto:2002yw})
\be
\phi(\vec{x}, \tau) \approx \tau^{\Delta_{\varphi}} \varphi (\vec{x}),\quad \mbox{with} \quad  \Delta_{\varphi} = \frac{3}{2}\left( 1- \sqrt{1-\frac{4m^2}{9H^2}}\right).
\ee
By demanding conformal invariance, it is possible to see that the field $\varphi(\vec{x})$  behaves as a conformal field with conformal weight $\Delta_{\varphi}$ \cite{Leblond2, Biagetti:2013qqa}. In general, under the assumption of slow roll inflation, it is possible to see that a scalar perturbation in de Sitter space behaves as a conformal field with conformal weight $\Delta_{\varphi} \sim {\cal O} (\epsilon, \eta)$, where $\epsilon$ and $\eta$ are the slow-roll parameters (see {\it e.g.} section 6.5 of \cite{Riotto:2002yw}).  \\
Now, we turn our attention to the vector field. The equation of motion for the vector derived from \eqref{coup} is
\be\label{eom1} 
\nabla_{\mu}\left[  f(\phi) \left( F^{\mu\nu} + \alpha \tilde{F}^{\mu\nu}  \right) \right]= 0.
\ee
Choosing the coupling function as $ f(\phi)= f(\tau)=(-H\tau)^{-2n}$ and the propagation along the $x$-axis, $\vec{k} = (k, 0,0)$, the above equation, in momentum space, is written as
\be \label{eomkalpha}
\left(\frac{\partial^{2}}{\partial \tau^{2}}    +  k^{2}  -  \frac{n (n + 1)}{\tau^2} + \lambda \frac{2\xi k}{\tau}  \right) {{w}_{\lambda}}(\tau, \vec{k}) = 0,
\ee 
where we have defined the canonical fields ${w}_\lambda \equiv \sqrt{f}{A}_\lambda$, $\lambda = \pm 1$ denote the helicity of the mode,  $\xi \equiv-n\alpha$,  and the transverse polarizations are defined such that $w_{\lambda} = {(w_y + \lambda i w_z)}/{\sqrt{2}}\,$. The solution of (\ref{eomkalpha}), in the region $|k\tau|\ll\xi\;,\xi \gg1$  can be well approximated by \cite{Caprini:2014mja,Almeida:2017lrq} 
\ba\label{approxA1}
&& {w}_{+}(k,\tau)\simeq \sqrt{-\frac{2\tau}{\pi}}e^{\frac{\pi \xi}{2}}K_{-(2n+1)}\left(\sqrt{-8 \xi k\tau}\right)\;,  \\
&& {w}_{-} (k,\tau)\simeq \sqrt{-\frac{2\tau}{\pi}}e^{-\frac{\pi \xi}{2}}Y_{-(2n+1)}\left(\sqrt{-8 \xi k\tau}\right)\;, 
\ea
where  $K_m(x)$ are the modified Bessel functions and $Y_{m} (x)$ are the Bessel functions of the second kind. The asymptotic solution, deep in the super horizon regime $|8\xi k \tau|\rightarrow 0$,   of those equations can be written as:
\be\label{apxmasymp}
{w}_{\lambda} (\tau, \vec{x}) \approx  {u}_{\lambda} (\vec{x})(-\xi H \tau)^{n+ 1} +  {v}_{\lambda} (\vec{x})(-\xi H \tau)^{-n}.
\ee
At this point, we notice that the dominant contribution of this solution depends on the value of the power of the coupling $n$. For $n>-1/2$, the mode $v$ dominates while, for $n<-1/2$, the mode $u$ dominates. In this work, we will use values close to $n=-2$, which reproduce a scale invariant spectrum for the sourcing modes, so, in the following, we will work only with the mode $u$.  {Consequently, in the super horizon limit, the field behaves like
\be 
\lim_{|\tau| \rightarrow 0} {w}_{\lambda} (\tau, \vec{x}) =  \tau^{n+ 1} {u}_{\lambda} (\vec{x}).  
\ee
Under dilatations, for instance, the coordinates change as $\tau'= \lambda \tau, \, x'_{\sigma} = \lambda x_{\sigma}$, and the vector \eqref{apxmasymp}  transforms as
\be\label{vectort} 
{w}_{\lambda}'   = \frac{\partial x_{\lambda}}{\partial x'_{\lambda'}} {w}_{\lambda'}= \lambda^{-1} {w}_{\lambda}, 
\ee
which, considering the dominant part of \eqref{apxmasymp} we deduce that
\be 
w_{\lambda} ' =\lambda^{n+1} \tau^{n+1} u_{\lambda}'(\vec{x}') =   \lambda^{-1}  \tau^{n+1} u_{\lambda}(\vec{x}),
\ee
where in the last equality we used the transformation law for a vector. This implies that the vector $u_{\lambda}(\vec{x})$ transforms as
\be 
u_{\lambda}'(\vec{x}')  = \lambda^{-(n+2)}u_{\lambda}(\vec{x}) 
\ee} 
Then, assuming conformal invariance  {of $u_{\lambda}(\vec{x})$ } in the super horizon regime, {this is $u'_{\lambda}(\vec{x})  = \lambda^{-\Delta_{u} }u_{\lambda}(\vec{x}) $ } (see {\it e.g.} \cite{Biagetti:2013qqa, Almeida:2017lrq}), we found that the conformal weight of the mode $u$ is:
\be\label{cw}
\Delta_{u} = n+2.
\ee
This is the conformal weight that we use for the vector contribution to the scalar and tensor perturbations in this paper. \\

Finally, to fix the nomenclature, we revisit here a few basics about the tensor modes in this setup (we use the notation used in, {\it e.g.}, \cite{Sorbo:2011rz,Bartolo:2015dga}). The equation for tensor modes comes from the metric 
\be\label{TPA}
ds^2=\frac{1}{(-H\tau)^2}\left[-d\tau^2+(\delta_{ij}+\gamma_{ij})dx^idx^j\right]\;,
\ee
with the gauge choice $\partial_i\gamma_{ij} = \gamma_{ii}=0$. From (\ref{TPA}), the  Einstein equations for the tensor modes are
\be\label{TMSol}
\gamma''_{ij}-\frac{2}{\tau}\gamma'_{ij}-\nabla^2 \gamma_{ij}=\frac{2}{M^2_P}\Pi_{ij}^{lm}T_{lm}^{EM}\;,
\ee
where $\Pi_{ij}^{lm}$ is a projector tensor and $T_{lm}^{EM}$ is the spatial part of the energy-momentum tensor:
\be
T_{lm}^{EM}=-f(\tau)a^{-2}A'_l A'_m+(\cdots)\;,
\ee
where the dots are other terms projected out by $\Pi_{lm}^{ij}$. 
Finally, we can write Eqs. (\ref{TMSol}) in terms of the helicity modes  $\gamma_\lambda(\tau,\vec{k})$ as
\be\label{eqtenmod}
\gamma''_\lambda-\frac{2}{\tau}\gamma'_\lambda-\nabla^2 \gamma_\lambda=\frac{2}{M^2_P}\Pi_\lambda^{lm}T_{lm}^{EM}\;, \quad \mbox{where} \quad \gamma_{\lambda} = \Pi^{ij}_{\lambda} \gamma_{ij}, \quad \mbox{and} \quad \Pi^{lm}_{\lambda}  \equiv \frac{\epsilon_{-\lambda}^l(\vec{k})\epsilon_{-\lambda}^m(\vec{k})}{\sqrt{2}}.
\ee
In the following we will reiteratively use the projector  $\Pi^{lm}_{\lambda}$ defined in \eqref{eqtenmod} to express the results in terms of the helicity components of the tensor modes $\gamma_{\lambda}$. Some useful properties of this projector are listed in Appendix \ref{AA2}. 
\subsection{Including the vector perturbations}
If we consider a perturbative contribution from the vector field to both tensor  and  primordial curvature perturbations, it means that, in the perturbative expansion, there will arise terms including the vector perturbations $\delta E_i (\vec{k})$. One can formally think about these corrections in the following form: the vector field contribute to the primordial curvature perturbation and to the tensor perturbation as follows
\be\label{zero1}
\zeta = \zeta^{(0)} + \zeta^{(1 )}, \qquad \gamma_{ij} = \gamma_{ij}^{(0)} + \gamma_{ij}^{(1)},
\ee
where the zero superscript $(0)$ stands for the vacuum contributions while the superscript $(1)$ stands for the ``sourced'' part\footnote{Henceforth we will call {\it sourced part} to the pure contributions of vector fields to the primordial perturbations.}. The perturbations come from the interaction Hamiltonian derived from the interaction term \eqref{coup}. In terms of the electric and magnetic fields:
\be
E_{i}= -\frac{\sqrt{f}}{a^2} A'_{i}(\tau) ,\qquad B_{i} =\frac{\sqrt{f}}{a^2} \epsilon_{ijk} \partial_{j} A_{k}(\tau),
\ee
the interaction Lagrangian is written as 
\be\label{LIEB}
{\cal L}_{I} = -\frac{1}{2} \left( \vec{B}^2 -  \vec{E}^2 \right)  + \alpha \vec{E}\cdot\vec{B},
\ee
and one derives the interaction Hamiltonian \cite{Bartolo:2015dga}
\be
{\cal H}_{I} = {\cal H}_1  + {\cal H}_g 
\ee
where
\ba\label{H1}
 {\cal H}_1 = -2n{a^4 }E^{(\rm{vev})}_i \int d^3k  \delta E_i (\vec{k}, \tau ) \zeta_{-\vec{k}}  \quad {\rm and}  \quad  {\cal H}_g =  {a^4 } E^{(\rm{vev})}_i \int d^3k  \delta E_j (\vec{k}, \tau ) \gamma_{ij,-\vec{k}}
\ea
are the leading contributions from the variation of the interaction Lagrangian \eqref{action0}. 
In the following, we only consider the linear order interaction Hamiltonian ${\cal H}_I$ to source  the scalar and tensor perturbations. A quadratic Hamiltonian term ${\cal H}_2$ will  also arise  but it will induce loop corrections that we shall not consider here. Schematically, the perturbation in the scalar and tensor perturbations will be sourced by the vector field through the interaction term \eqref{LIEB} as follows  
\be\label{pertdA}
 \zeta^{(1)} \propto  (\hat{E} \cdot \delta \vec{E}  )  \zeta^{(0)}, \qquad  \gamma_{\lambda}^{(1)}\propto (\hat{E} \cdot \delta \vec{E}) \gamma_{\lambda}^{(0)},
\ee
here, $\hat{E}_i = E_{i}/(E_{k}E_{k})^{1/2}$  is the unit vector along $E^{(\rm{vev})}_i$. This implies that the sourced tensor perturbation $ \gamma_{ij}^{(1)}$ is introduced as follows
\be\label{pertdA2}
 \gamma_{ij}^{(1)}\propto (\hat{E} \cdot \delta \vec{E})  \gamma_{ij}^{(0)}.
\ee
The way in which the vector perturbations enters into the calculation of scalar and tensor perturbations is detailed in Refs. \cite{Bartolo:2012sd,Bartolo:2014hwa,Bartolo:2015dga} through the use of the ``in-in'' formalism. Eqs. \eqref{pertdA} and \eqref{pertdA2} are not meant to be precise equalities, they are only a guide for tracking the presence of vector perturbations in the calculations.  For our purpose here, it is enough to track the presence of a vector perturbation $\delta E_{i}$ carrying an object with an index which transforms as a $O(3)$ vector. \\
Eqs. \eqref{pertdA} and \eqref{pertdA2} are the expressions that define the structure of the perturbations that we will consider in this work. With them, we will derive the structure of the two and three point correlation functions consistent with three dimensional conformal symmetry generated by such interactions. Due to the particular form \eqref{pertdA} of introducing the perturbations, we assume that the sourced perturbations, in the super horizon limit, behave like a conformal primary field with conformal weight 
\be\label{cwpp}
\Delta_{\zeta1} = \Delta_{\zeta0} + \Delta_{u}, \qquad \mbox{and} \qquad   \Delta_{\gamma1} = \Delta_{\gamma0} + \Delta_{u},
\ee
where  $\Delta_{\zeta0}, \Delta_{\gamma0} \sim {\cal O}(\epsilon, \eta)$ comes from the contribution  of the scalar field $H \delta \varphi /\dot{\varphi}$ to the primordial curvature perturbation and  $\Delta_{u}$ is the conformal weight of the source vector field obtained in \eqref{cw}.

\section{The spectrum} \label{spectrum}
Here we study the structure of the two point functions involving the primordial curvature perturbations and the tensor perturbations, this is, we consider the symmetry constraints on the following two point functions:
\ba
\langle \zeta \zeta \rangle, \quad \langle \zeta \gamma \rangle, \quad \langle \gamma \gamma \rangle.
\ea
First, we write a general form of the two point correlator restricted by momentum conservation, reality condition of the correlations, and gauge invariance implementing a particular gauge choice. This results in the conditions for the vanishing of the trace and the divergence of the vector and tensor perturbations. Afterwards, we use Ward identities to determine the scaling behaviour of the correlators \cite{Biagetti:2013qqa, Almeida:2017lrq}. To this end, and for the two point correlators, it is enough to use the Ward identity related with dilatations.  {For the two point function, the Ward identity related with special conformal transformations (SCT) implies the vanishing of correlators involving mixed fields with different scaling dimension (see {\it e.g } the section 1.3 of \cite{Ginsparg:1988ui}). Aside from that, the SCT results are redundant. The scaling and the structure of non-vanishing two point correlators can be derived solely from the Ward identity for the dilatations and from the transverse and trace free gauge choice for the perturbations. } 
 {As mentioned in the introduction, Ref. \cite{Bzowski:2013sza} contains a vast list of results for the correlators of conformal fields in momentum space. It results useful to relate our notations with theirs in order to campare the results. For the scalar ${\cal O} $, the vector $J_l$ and the tensor field $T_{ij}$ in \cite{Bzowski:2013sza}, we will use the following connection through the text $\zeta^{(0)}\sim {\cal O}$, $\delta E_{l} \sim J_{l}$ and $\gamma^{(0)}_{ij}\sim T_{ij}$.  }

\subsection{Scalar spectrum } \label{z2p}
The scalar spectrum comes from the following contributions:
\be
\langle \zeta \zeta \rangle = \langle \zeta^{(0)} \zeta^{(0)} \rangle + \langle \zeta^{(0)} \zeta^{(1)} \rangle  + \langle \zeta^{(1)} \zeta^{(0)} \rangle  + \langle \zeta^{(1)} \zeta^{(1)} \rangle,
\ee
where the terms $\langle \zeta^{(0)} \zeta^{(1)} \rangle$ and $ \langle \zeta^{(1)} \zeta^{(0)} \rangle$, according with \eqref{pertdA}, are proportional to $  \hat{E}_{k} \langle  \zeta^{(0)}  \delta E_{k} \rangle$.  { This function is related with $\langle J_i \mathcal{O} \rangle$ in \cite{Bzowski:2013sza} which, as mentioned before, is zero as a consequence of symmetry under SCT.}  
 {We can see that explicitly in the case of the vector perturbations considered here if we construct a two point function with a single vector index. In general, given that the vector perturbation is imposed to be divergence free, this correlator can be constructed as a combination of polarization vectors carrying a single vector index (some relevant properties of polarization vectors and tensors are listed in Appendix B). The only option for such function, with a single vector index, satisfying simultaneously the dilatations and SCT Ward identities is to be zero.  }
In consequence, for the scalar spectrum including the effect of the vector sources, we are left with 
\be
\langle \zeta \zeta \rangle = \langle \zeta^{(0)} \zeta^{(0)} \rangle + \langle \zeta^{(1)} \zeta^{(1)} \rangle.
\ee
To compute the vacuum scalar spectrum 
we propose a general form which, in this case, is written as 
\begin{equation}
\langle \zeta^{(0)}(\vec{k}_1)\zeta^{(0)}(\vec{k}_2)\rangle=\delta(\vec{k}_{12}) P_0(k_1),
\end{equation}
where {$\vec{k}_{12\cdots n} = \vec{k}_{1} + \cdots + \vec{k}_{n}$}. Now, we use the Ward identity for dilatations given by (see Appendix \ref{fm} for the explicit form of these identities)
\begin{equation}
\left[-3+2\Delta_{\zeta0}-k_{1l} \frac{\partial}{\partial k_{1l}} \right] \langle \zeta^{(0)}(\vec{k}_1) \zeta^{(0)}(\vec{k}_2) \rangle'=0,
\end{equation}    
where the apostrophe in the correlation function means that  we factor out the delta function. The solution of this equation is $  P_0(k_1)=\alpha_0k_1^{-3+2\Delta_{\zeta_0}}, $  then, the vacuum  spectrum is written as
\begin{equation}
\langle \zeta^{(0)}(k_1)\zeta^{(0)}(-k_1)\rangle' = \alpha_0 k^{-3+2\Delta_{\zeta_{0}}}.
\end{equation}
On the other hand, for the  scalar spectrum sourced by vector fields, we  propose the following form  
\begin{equation}
\langle \zeta^{(1)}(k_1)\zeta^{(1)}(-k_1)\rangle'= \hat{E}^l\hat{E}^m P_{lm}(\vec{k}_1),
\end{equation}
where  $P_{lm}$ is constructed out of the $O(3)$ invariant tensors $\delta_{lm}, k_{l}, k_{m}$ and the Levi-Civita tensor $\eta_{lmn}$. Due to gauge invariance, zero divergence is imposed over the indices $l$ and $m$. The general object that result from these conditions is
\begin{equation}
P_{lm}=P_1(k_1)\Delta_{lm}+P_2(k_1)\hat{\eta}_{lm}, \quad \mbox{where} \quad \Delta_{ij}=\delta_{ij}-\hat{k}_{1i}\hat{k}_{1j}  \quad \mbox{and} \quad \hat{\eta}_{lm}=\eta_{lma} \hat{k}_1^a.
\end{equation}
Reality of this correlator implies that $P_1(k_1)$ is real and $P_2(k_1)$  is pure imaginary. The Ward identity for dilatations implies the scaling of the functions $P_{1}$ and $P_{2}$:
\begin{align}
P_1(k_1)=\alpha_1 k_1^{-3+2\Delta_{\zeta1}} , \quad
P_2(k_1)=i \alpha_2 k_1^{-3+2\Delta_{\zeta1}},
\end{align}
where $\alpha_i$ are real constants and $\Delta_{\zeta1}= \Delta_{\zeta0} + \Delta_{u}$, as obtained in \eqref{cwpp}, is the conformal weight of the sourced perturbations. As a result, the scalar two-point correlation function is written as 
\begin{equation}
\langle \zeta^{(1)}(k_1)\zeta^{(1)}(-k_1)\rangle' = \alpha_1  k_1^{-3+2\Delta_{\zeta0} + 2\Delta_{u}} \hat{E}^l\hat{E}^m \Delta_{lm},
\end{equation}
where we can notice the presence of a quadrupolar term due to the vector perturbations. This quadrupolar structure has been extensively studied in the context of vector field models of inflation, see e.g. \cite{Bartolo:2014hwa,Bartolo:2015dga, Almeida:2017lrq}. We can notice here that, despite of the parity breaking nature of the system, there are no parity breaking signatures appearing in the scalar spectrum. \\
Adding the vacuum and the sourced parts, the total scalar spectrum obtained is: 
\ba \nonumber
\langle \zeta(\vec{k}_{1}) \zeta(-\vec{k}_{1})  \rangle' &=& \langle \zeta^{(0)}(\vec{k}_{1}) \zeta^{(0)}(-\vec{k}_{1})  \rangle' + \langle \zeta^{(1)}(\vec{k}_{1}) \zeta^{(1)}(-\vec{k}_{1})  \rangle' \\
&= &  \alpha_0 k_{1}^{-3+2\Delta_{\zeta_0}} \left[ 1 + g_{\zeta} (k_{1}) \left( 1 - (\hat{E}\cdot \hat{k}_1)^2 \right) \right],\quad \mbox {where} \quad  g_{\zeta} (k_{1}) \equiv \left( \frac{\alpha_1 }{\alpha_0} \right)  k_{1}^{2\Delta_{u} }. 
\ea 
It is interesting to remark that the parameter $g_{\zeta}$ acquires a  scaling of the form $g_{\zeta} \sim k_{1}^{2\Delta_{u} } = k_{1}^{4 + 2n}$ as a consequence of the sourcing of the vector modes. In the case $n=-2$, the parameter $g_\zeta$ is scale invariant, but any deviation  from this value can introduce a significant scale dependence due to this sourcing mechanism. 
\subsection{The tensor-scalar spectrum} \label{gza}
Here we consider the mixed tensor-scalar two point function, which is written as  
\be
\langle \gamma_{ij} \zeta \rangle = \langle \gamma_{ij}^{(0)} \zeta^{(0)} \rangle + \langle \gamma_{ij}^{(0)} \zeta^{(1)} \rangle  + \langle \gamma_{ij}^{(1)} \zeta^{(0)} \rangle  + \langle \gamma_{ij}^{(1)} \zeta^{(1)} \rangle.
\ee
To evaluate the zero-th order contribution to this correlator we use the divergence {and trace} free condition of the tensor perturbations $k_{1}^i\gamma_{ij}(\vec{k}_1) {=\gamma_{ii}(\vec{k}_1)}=0$. For the same reason mentioned before, and given that $\langle \gamma_{ij}^{(0)} \zeta^{(0)} \rangle$ involves fields with different scaling dimension, this correlator is zero as a consequence of invariance under SCT. 
This implies that the vacuum scalar and tensor perturbations  are not correlated and it is remarkable that this result can be derived only from symmetry arguments. The same happens for the terms $\langle \gamma_{ij}^{(1)} \zeta^{(0)} \rangle $ and $\langle \gamma_{ij}^{(0)} \zeta^{(1)} \rangle$, they are zero due to invariance under SCT. 
Then,  we are left with
\be
\langle \gamma_{ij} \zeta \rangle = 
\langle \gamma_{ij}^{(1)} \zeta^{(1)} \rangle,
\ee
which means that the scalar-tensor two point function acquires a non zero value only due to the presence of the sourced primordial perturbations. Considering the structure of the interaction terms, which at leading order include $\hat{E}_i$ and $\delta E_j$, we write a general expression for the correlator in the from 
\be\label{gammaAA}
\langle \gamma_{ij}^{(1)} (\vec{k}_1) \zeta^{(1)} (\vec{k}_2) \rangle= \delta(\vec{k}_{12})\hat{E}^l\hat{E}^m{\rm B}_{ijlm}, 
\ee
where ${\rm B}_{ijlm}$ is the more general $O(3)$ invariant tensor with four indices, constructed out of the $O(3)$ invariant objects $k_{i}, \delta_{ij}, \eta_{ijk}$. In the following we will write ${\rm B}_{ijlm}$ as a function of only $\vec{k}_1$ due to momentum conservation, while the dependence on $\vec{k}_2$ is contained in the delta function $\delta(\vec{k}_{12})$. The tensor ${\rm B}_{ijlm}$ can be decomposed in a basis of $O(3)$ invariant tensors  
\be
{\rm B}_{ijlm}= \sum_{n=1}^{4} b_n B_{ijlm}^{(n)}\;,
\ee 
where $b_n$ are functions of the momenta, and the different, independent objects $B^{(n)}_{ijlm}$\footnote{ It is important to mention that the superscript $(n)$ here is not related with the order of the perturbation expansion as introduced in \eqref{zero1}. We use the same notation but the meaning should be understood from the context.} obey traceless, divergenceless conditions and are symmetric in the indices $(ij)$,  since the left hand side of (\ref{gammaAA}) is symmetric in those indices. The previous conditions are satisfied by the next set of tensors (we give some details about the derivation of these tensors in Appendix \ref{umatrices})
\begin{align}\label{Bijred1}
B^{(1)}_{ijlm}&=\Delta_{mj}\Delta_{il}+\Delta_{lj}\Delta_{im} -\Delta_{ij}\Delta_{lm}, \\ \label{Bijred2}
B^{(2)}_{ijlm}&=\Delta_{il}\hat{\eta}_{jm}  + \Delta_{lj}\hat{\eta}_{im}  - \Delta_{ij}\hat{\eta}_{lm}   ,\\ \label{Bijred3}
B^{(3)}_{ijlm}&=\hat{\eta}_{il} \hat{\eta}_{jm}  + \hat{\eta}_{im}  \hat{\eta}_{jl}  -\Delta_{ij}\Delta_{lm},\\ \label{Bijred4}
B^{(4)}_{ijlm}&=\Delta_{im} \hat{\eta}_{jl}  + \Delta_{mj} \hat{\eta}_{il}  + \Delta_{ij} \hat{\eta}_{lm}.
\end{align}
From the list before, we can notice that $B^{(1)}_{ijlm}$  and $B^{(3)}_{ijlm}$ are symmetric in the indices $(lm)$ and that $B^{(2)}_{ijml}=B^{(4)}_{ijlm}$. Due to  this relation 
 and taking into account that the contraction with $\hat{E}^{l}\hat{E}^{m}$ don't recognize the difference between $l$ and $m$, we will use only the tensor $B^{(2)}_{ijlm}$ for the evaluation of this  correlator. Moreover, by using identities of the Levi-Civita symbol, it is possible to realize that $B^{(3)}_{ijlm} = -B^{(1)}_{ijlm}$ (for details, see Appendix \ref{A11}). We can symmetrize the resulting two tensors defining 
 \begin{align}\label{Tijred1}
T^{(1)}_{ijlm}&=\frac{1}{2}B_{ijlm}^{(1)},\\ \label{Tijred2}
T^{(2)}_{ijlm}&=\frac{i}{4}(B_{ijlm}^{(2)}+B_{ijlm}^{(4)}), 
\end{align}
as the independent symmetric objects in the indices $(ij),\, (lm)$,  consistent with  trace and divergence free conditions.
Additionally, we can use the reality of the correlator, $\langle  \gamma_{ij}^{(1)}(\vec{k}_1) \zeta^{(1)} (-\vec{k}_1)  \rangle^{\dagger} = \langle \gamma_{ij}^{(1)}(\vec{k}_1) \zeta^{(1)} (-\vec{k}_1) \rangle$, to impose conditions over the functions $b_n$, this implies 
\be
 \sum_{n=1}^{2} b_n^*(k_1) T_{ijlm}^{(n)}(-\vec{k}_1)=\sum_{n=1}^{2} b_n(k_1) T_{ijlm}^{(n)}(\vec{k}_1)\;,
\ee
and therefore $b_1^*= b_1$, $b_2^*=-b_2$,  which means that $b_1$ is a real function, while $b_2$ is a pure imaginary quantity. 
In this form, the correlation function can be written as 
\be
\langle \gamma_{ij}^{(1)}(\vec{k}_1)  \zeta^{(1)} (-\vec{k}_1) \rangle=\delta(\vec{k}_{12})\hat{E}^l\hat{E}^m\left[ b_1(k_1) T_{ijlm}^{(1)}(\vec{k}_1) + i   b_2(k_1) T_{ijlm}^{(2)}(\vec{k}_1) \right]\;.
\ee 
Finally, we apply the Ward identity for dilatations over the last expression of the correlator: 
\be
\left[-3+\Delta_{\gamma0}+\Delta_{\zeta0} + 2\Delta_{u} -k_{1l} \frac{\partial}{\partial k_{1l}} \right] \langle \gamma_{ij}^{(1)}(\vec{k}_1) \zeta^{(1)}(-\vec{k}_1) \rangle'=0\;.
\ee
Given that  $k_{1a} \frac{\partial}{\partial k_{1a}}T_{ijlm}^{(n)}=0$, we obtain an equation for the functions $b_n$
\be\label{Dwigz}
\left[-3+\Delta_{\gamma0}+\Delta_{\zeta0} + 2\Delta_{u}-k_{1l} \frac{\partial}{\partial k_{1l}} \right]b_n(k_1)=0\;,
\ee
whose solution is
\be
b_n(k_{1})=B_nk_{1}^{-3+\Delta_{\gamma0}+\Delta_{\zeta0} + 2\Delta_{u}}\;,
\ee
where $B_n$ are constants. Finally, by using  the identities of the projector $\Pi^{lm}_{\lambda}$ listed in \ref{AI}, we project into the helicity basis obtaining:
\be
\langle \gamma_{\lambda}^{(1)}(\vec{k}_{1}) \zeta^{(1)}  (-\vec{k}_{1}) \rangle' =   k_{1}^{-3 + \Delta_{\gamma0}+\Delta_{\zeta0} + 2\Delta_{u} } B_\lambda  \hat{E}^l \hat{E}^m  \Pi^{lm}_{\lambda},
\quad {\rm where} \quad B_\lambda=B_1 +\lambda B_2.
\ee
It is remarkable that this result shows an explicit dependence on the helicity in the coefficient $B_\lambda$ coming from the tensor $T^{(2)}_{ijlm}$.  Given that the contribution from the vacuum perturbations in this case is null, we get that the non zero value of this correlation function is an effect coming purely from the vector fields sources, then: 
\be
\langle \gamma_{\lambda}(\vec{k}_{1}) \zeta  (-\vec{k}_{1}) \rangle ' = \langle \gamma_{\lambda}^{(1)}(\vec{k}_{1}) \zeta^{(1)}  (-\vec{k}_{1}) \rangle' =   k_{1}^{-3+\Delta_{\gamma0}+\Delta_{\zeta0} + 2\Delta_{u} } B_\lambda  \hat{E}^l \hat{E}^m  \Pi^{lm}_{\lambda}.
\ee
This result, coincides, in structure, with the results reported in \cite{Bartolo:2015dga} and with  \cite{Ohashi:2013qba} for the parity conserving limit $\xi \rightarrow 0$ of this model, but, as we mentioned in the introduction, it is necessary to emphasize that with this approach we are not able to obtain the amplitudes of the correlators, so, the function $B_{\lambda}$ is not fully determined by this method. However, it is interesting to consider some limit cases of this function.  For instance, when  $B_1\approx B_2$  we obtain $B_{+} \gg B_{-},$ and we see a clear difference in the correlation of the different polarizations. This limit exhibits the parity violating nature of the system. On the other hand, the limit $B_2\approx 0$ leads to a parity conserving situation in which $B_{+} \approx B_{-}$. Another limit, $B_1\approx 0$ leads to the case $B_{+} \approx -B_{-}$ in which the correlations have opposite sign.

\subsection{Tensor tensor spectrum } \label{ggv}
Finally, we compute the $\langle\gamma \gamma\rangle$ correlator. As before, we separate the vacuum and the sourced contributions in the following way:
\be 
\langle    \gamma_{ij}  \gamma_{kl}  \rangle = \langle   \gamma_{ij}^{(0)}  \gamma_{kl}^{(0)}   \rangle  +  \langle    \gamma_{ij}^{(0)}  \gamma_{kl}^{(1)}   \rangle 
+  \langle    \gamma_{ij}^{(1)}  \gamma_{kl}^{(1)}   \rangle,
\ee
however, for the same reasons mentioned before, the mixed correlator $ \langle    \gamma_{ij}^{(0)}  \gamma_{kl}^{(1)}   \rangle$ is equal to zero. For the vacuum part we can write a general expression in terms of  \eqref{Tijred1} and \eqref{Tijred2}
\be
\langle \gamma_{ij}^{({\rm 0})}(\vec{k}_1) \gamma_{lm}^{({\rm 0})}(\vec{k}_2) \rangle= \delta(\vec{k}_{12})\sum_{n=1}^{2} t_n(k_1) T_{ijlm}^{(n)}\;,
\ee 
where, $t_{n}(k_1) $ are arbitrary functions of the momentum. This correlator is constructed with the objects $k_{i}, \delta_{ij}, \eta_{ijk}$, which are invariant under $O(3)$, and fulfill the  traceless and divergence free conditions, and is symmetric under $i\leftrightarrow j$, $l\leftrightarrow m$. The correlator must be real: $\langle \gamma_{lm}^{({\rm 0})}(\vec{k}) \gamma_{ij}^{({\rm 0})}(-\vec{k}) \rangle^\dagger=\langle \gamma_{ij}^{({\rm 0})}(\vec{k}) \gamma_{lm}^{({\rm 0})}(-\vec{k}) \rangle$, which implies 
that $t_1$ is real and $t_2$ is a pure imaginary. 
Again, by using the Ward identity for dilatation over this correlator and conclude that   
\be
t_n(k_1)=T_n k_{1}^{-3+2\Delta_{\gamma_0}}\;,
\ee
and contracting with the projector  $\Pi^{ij}_{\lambda}$ we obtain:
\be
\langle \gamma_\lambda^{({\rm 0})}(\vec{k}_1) \gamma_{\lambda'}^{({\rm 0})}(-\vec{k}_1) \rangle = \delta(\vec{k}_{12})\Pi_{\lambda}^{ij}\Pi_{-\lambda'}^{lm}  \langle \gamma_{ij}^{({\rm 0})}(\vec{k}_1) \gamma_{lm}^{({\rm 0})}(-\vec{k}_1) \rangle',
\ee  
where the $-\lambda'$ comes from momentum conservation given that the Dirac's $\delta(\vec{k}_{12})$ implies that the second projector acts over $-\vec{k}_{1}$. Finally, by using the identities \eqref{pt123},  the result is
\be
\langle \gamma_{\lambda}^{({\rm 0})}(\vec{k}_1) \gamma_{\lambda'}^{({\rm 0})}(-\vec{k}_1) \rangle' =   k_{1}^{-3+2\Delta_{\gamma_0}} T_\lambda \delta_{\lambda, \lambda'}, \quad {\rm where} \quad T_\lambda=T_1 + \lambda T_2. 
\ee  
Again, one can see that the result depends on the helicity. There is no correlation between different tensor helicities, and the spectra for both helicities are different. An estimate of the difference between the two spectra was obtained in a pseudoscalar related model in  \cite{Sorbo:2011rz}. In this reference, it was estimated that the difference in the amplitudes between both spectra $T_{\pm}$ is about two orders of magnitude. With the symmetry based approach that we follow here, we can't do any estimate about this difference. We can get, however, as in the case of the correlator $\langle \gamma \zeta \rangle$, limit cases in which this difference is noticeable.  For instance, the appropriate limit to see a noticeable difference is $T_1\approx T_2$. In this case we get $T_{+}\gg T_{-}$.  \\ 

Now, we move to calculate the vector field source contribution, which can be written as
\be \label{ggaa}
\langle \gamma_{ij}^{(1)}(\vec{k}_1) \gamma_{lm}^{(1)}(\vec{k}_2) \rangle = \delta(\vec{k}_{12})   \hat{E}^k \hat{E}^n  {\rm U}_{ijklmn},
\ee
where ${\rm U}_{ijlmkn}$ is an object, which is symmetric and trace free in the indices $i \leftrightarrow j$ and $l \leftrightarrow m$ and divergenceless in all of them. ${\rm U}_{ijlmkn}$ can be constructed by using the objects: $B_{ijlm}^{(s)}$, $\Delta_{ij}$ and $\hat\eta_{ij}$ as shown in appendix \ref{umatrices}. All in all we have,
\be
\langle \gamma_{ij}^{(1)}(\vec{k}_{1}) \gamma_{lm}^{(1)}  (-\vec{k}_{1}) \rangle= \delta(\vec{k}_{12}) \hat{E}^k \hat{E}^n \sum_{n=1}^{16} u_n(k_1) U_{ijklmn}^{(n)}\;,
\ee
but, inspecting the explicit form of the tensors $U^{(n)}$ in appendix \ref{umatrices}, and considering that we contract with the symmetric pair $ \hat{E}^k \hat{E}^n$, we reduce the number of necessary tensors to the ones listed in \eqref{P14}, \eqref{P25} and \eqref{P36}.   By imposing the reality of the correlator we obtain 
\begin{equation}
\langle \gamma_{ij}^{(1)}(\vec{k}_{1})  \gamma_{lm}^{(1)}(-\vec{k}_{1})  \rangle= \delta(\vec{k}_{12})\hat{E}^k \hat{E}^n\left[\sum_{s={1}}^{3} u_s(k_1)P_{ijklmn}^{(s)}+ i \sum_{r=4}^{6}  u_{r}(k_1)P_{ijklmn}^{(r)}\right]\;,
\end{equation}
where $u_{i}$ are real functions. Finally, by using the Ward identity for dilatation and noticing that the tensors $P^{(n)}_{ijlmkn}$ are not affected by the operator $k_{1l} \frac{\partial}{\partial k_{1l}}$, we obtain
 \begin{equation}
u_{n}(k_{1})=U_n k_{1}^{-3+2\Delta_{\gamma_1}},
\end{equation}
where $U_{n}$ are real constants and $\Delta_{\gamma_1} = \Delta_{\gamma0} + \Delta_{u}$. Then, we have
\be
\langle \gamma_{ij}^{(1)}(\vec{k}_{1})  \gamma_{lm}^{(1)}(-\vec{k}_{1}) \rangle= \delta(\vec{k}_{12})k_1^{-3+2\Delta_{\gamma{0}} + 2\Delta_{u} }\hat{E}^k \hat{E}^n\left[\sum_{s=1}^{3}U_s P_{ijklmn}^{(s)}+i\sum_{r=4}^{6} U_{r} P_{ijklmn}^{(r)} \right]\;.
\ee
Contracting with the projectors to the helicity basis, we get 
\be
\langle \gamma_{\lambda}^{(1)}(\vec{k}_{1}) \gamma_{\lambda'} ^{(1)}(-\vec{k}_{1})  \rangle'= \Pi_{\lambda}^{ij}\Pi_{-\lambda'}^{lm}\langle \gamma_{ij}^{(1 )}(\vec{k}_{1})  \gamma_{lm}^{(1)}(-\vec{k}_{1})\rangle \propto k_1^{-3+2\Delta_{{\gamma0 } } + 2\Delta_{u} } \delta_{\lambda,\lambda'}\hat{E}^k\hat{E}^n\Delta_{kn}U_\lambda ,
\ee
where we have used the identities \eqref{Piida}-\eqref{eepipi}, we renamed the constants $U_s$ and 
\be
{U_{\lambda}  \equiv U + \lambda V = \frac{1}{2}\left[ U_1+U_2 +U_3 +\lambda \left(U_{4} + U_{5}+U_{6}   \right) \right]  }.
\ee
To summarize, symmetry under conformal transformations, restricts the structure of the tensor-tensor spectrum to be as follows:
\ba \nonumber
\langle \gamma_{\lambda}(\vec{k}_{1}) \gamma_{\lambda'}(-\vec{k}_{1})  \rangle' &=& \langle \gamma_{\lambda}^{(0)}(\vec{k}_{1}) \gamma_{\lambda'} ^{(0)}(-\vec{k}_{1})  \rangle' + \langle \gamma_{\lambda}^{(1)}(\vec{k}_{1}) \gamma_{\lambda'} ^{(1)}(-\vec{k}_{1})  \rangle' \\
&= & T_{\lambda} k_1^{-3+2\Delta_{\gamma{0}}   } \delta_{\lambda,\lambda'} \left( 1 + g_{\gamma} (k_{1}) \hat{E}^k\hat{E}^n\Delta_{kn} \right), \quad \mbox{where} \quad  g_{\gamma} (k_{1}) \equiv \left( \frac{U_{\lambda}}{T_{\lambda}} \right)  k_{1}^{2\Delta_{u} }. \qquad
\ea 
 We notice again, that, as in the scalar-scalar case, the structure of the correlation function exhibits a quadrupole term. This quadrupolar structure was already derived in \cite{Bartolo:2015dga} using the in-in formalism. As before, the correlation functions for the 2 polarizations are different and the correlation between  two different polarizations is zero. 
\section{The bispectrum} \label{bispectrum}
In this section, we use the approximate conformal symmetry in the inflationary regime to elucidate the structure of the three-point functions in momentum space. In this case, the Ward identity for dilatations is not enough to determine the form of the correlation functions, since these depend on the three momenta, $\vec{k}_1, {\vec{k}_2}$ and ${\vec{k}_3}$, and therefore we need  additional equations to obtain the dependence with all the three momenta. So, it is necessary to use the Ward identities for  SCT to determine the structure of the correlator. A complete and very detailed analysis of the restrictions imposed by SCT in momentum space can be found in \cite{Maldacena:2011nz, Coriano:2013jba} for scalar correlators and  in \cite{Bzowski:2013sza} for correlators of scalar, vector and tensor operators. In this section, we rely on the results found in \cite{Bzowski:2013sza} for parity invariant correlations and extrapolate their results to include parity breaking. 
As mentioned in the introduction, this is not a trivial task, because to extrapolate the results from one context to the other, we need to obtain the appropriate tensor decomposition including the parity breaking contributions. To this end, we use the objects $T_{ijlm}^{(a)}(\vec{k}_p)$, $\Delta_{ij}(\vec{k}_p)$ and $\hat{\eta}_{ij}(\vec{k}_p)$ with $p=1,2,3$ introduced in section \ref{spectrum}.  \\
In reference \cite{Bzowski:2013sza}, the Ward identities related with SCT are solved  in terms of a triple Bessel $K$ functions integral 
\be\label{tripleK}
 J_{N\{p_j\}} (k_1, k_2, k_3) =\int_0^\infty dx x^{1/2+N}\prod_{j=1}^3k_j^{\Delta_j-3/2+p_j}K_{\Delta_j-3/2+{p_j}}(k_jx)\;, 
\ee
where the $K_n(x)$ are the modified Bessel functions of the second kind, $\Delta_j$ is the conformal weight of the operator, $k_j$ is the magnitude of the corresponding momenta and $N$ is the number of momenta contracted with the tensors  $T_{ijlm}^{(1)}$ and $\Delta_{ij}$. 
Furthermore, we can see that for the primary SCT Ward identities, defined in (\ref{psctwi}), the $J_{N\{p_j\}}$ satisfies
\begin{equation}
\mathcal{K}_{ab}J_{N\{p_j\}}=2p_aJ_{N+1\{p_j-\delta_{aj}\}}-2p_b J_{N+1\{p_j-\delta_{bj}\}}.
\end{equation} 
 In these equations, $p_j$ is an integer that gives the freedom of making the primary SCT Ward identities homogeneous or not, according to the particular case. We revisit briefly some details about the primary conformal equations in Appendix \ref{fm}. For more details see \cite{Bzowski:2013sza}.
\subsection{Scalar Bispectrum }
Next, we compute the bispectrum including the effects from the  vector field. Similar to what we write in the previous section for the power spectrum, we can introduce the contributions sourced by the vector field as follows:
\ba\label{generalSB}
\langle \zeta \zeta \zeta \rangle &=& \langle \zeta^{(0)} \zeta^{(0)}  \zeta^{(0)}  \rangle + \langle \zeta^{(1)} \zeta^{(0)}  \zeta^{(0)} \rangle + (\mbox{2 perms.}) 
+ \langle \zeta^{(1)}  \zeta^{(1)} \zeta^{(0)} \rangle  + (\mbox{2 perms.}). 
\ea
In the following we will calculate every piece separately. To start, the vacuum bispectrum can be written as 
\be
\langle \zeta^{(0)}(k_1)\zeta^{(0)}(k_2)\zeta^{(0)}(k_3)\rangle=\delta(k_{123}) F(k_1,k_2,k_3),
\ee
where $F$ is a scalar function of the momenta consistent with conformal symmetry. The solution for the Ward identities for conformal symmetries in momentum space  found in  \cite{Coriano:2013jba, Bzowski:2013sza} is
\ba\label{z0z0z0}
F(k_1,k_2,k_3)&=&  a_0 J_{0\{000\}} \\ \nonumber
&=& a_0 \int_0^\infty dx x^{1/2} (k_1k_2 k_3)^{\Delta_{\zeta0} -3/2}K_{\Delta_{\zeta1}-3/2}(k_1 x) K_{\Delta_{\zeta0}-3/2}(k_2x) K_{\Delta_{\zeta0}-3/2}(k_3 x),
\ea
where $a_0$ is a normalization constant and we have used the notation for the triple-K integrals in \eqref{tripleK}. \\
Now, we move to compute  
the second piece of (\ref{generalSB}), 
which is constructed by considering the most general object that is divergenceless with respect to $k_1$. This can be written as a linear combination of  $\Delta_{la}({k}_1)$ and $\hat{\eta}_{la}({k}_1)=\eta_{lac}\hat{k}_{1c}$ as follows:
\ba\label{z1z0z0}
\langle \zeta^{(1)}(k_1)\zeta^{(0)}(k_2)\zeta^{(0)}(k_3)\rangle= \delta(k_{123}) \hat{E}^l 
 \left[\Delta_{la}(k_1)+i \alpha_1 \hat{\eta}_{la}(k_1)  \right]A^{a},
\ea
where the function
\be 
A^{a} = A_1(k_1, k_2, k_3) k_{2}^{a}, 
\ee
is a general function which doesn't vanish upon contraction with  $\Delta_{la}({k}_1)$ and $\hat\eta_{la}({k}_1)$. The function $A^{a}$ is symmetric under the change $k_2 \leftrightarrow k_3$.
This is the general form found in \cite{Bzowski:2013sza} for this correlator, the only difference here is that we also consider the odd part $\hat{\eta}_{la}(k_1)$. The function $A_1(k_1, k_2, k_3)$ solves the primary conformal Ward identities and is expressed in terms of the integrals \eqref{tripleK} as  
\ba
A_1(k_1, k_2, k_3) &=& a_1 J_{1\{000\}} \\ \nonumber
&=& a_1  \int_0^\infty dx x^{3/2} k_1^{\Delta_{\zeta1} -3/2} (k_2 k_3)^{\Delta_{\zeta0} -3/2}K_{\Delta_{\zeta1}-3/2}(k_1 x) K_{\Delta_{\zeta0}-3/2}(k_2x) K_{\Delta_{\zeta0}-3/2}(k_3 x),
\ea
where $a_1$ is a normalization constant. 
With the previous results, the total contribution from first order sourced vector perturbations becomes
\ba\label{z1z0z0}
\langle \zeta^{(1)}(k_1)\zeta^{(0)}(k_2)\zeta^{(0)}(k_3)\rangle &+&( \mbox{2 perms.})  \\ \nonumber
 &=& \delta(k_{123}) a_1 k_2 J_{1\{000\}}  
 \left[  \hat{E} \cdot \hat{k}_2 -   (\hat{E} \cdot \hat{k}_1 )  (\hat{k}_1 \cdot \hat{k}_2 ) + i \alpha_1 \hat{E}\cdot (\hat{k}_2 \times \hat{k}_1)\right]+ (\mbox{2 perms.}).
\ea
Finally, we compute the second order vector sourced part $ \langle \zeta^{(1)}  \zeta^{(1)} \zeta^{(0)} \rangle $. This contribution is constructed out with the appropriate combinations of $\Delta_{ab}(k_p)$ and $\hat{\eta}_{ab}(k_p)$ for the momenta $\vec{k}_p $ ($p=1,2$): 
\ba\label{z1z1z0de}
&&\langle \zeta^{(1)}(k_1)\zeta^{(1)}(k_2)\zeta^{(0)}(k_3)\rangle = \delta(k_{123}) \hat{E}^l \hat{E}^m\\ \nonumber
&&\qquad\qquad \times \left[\Delta_{la}(k_1)\Delta_{mb}(k_2) +i\alpha_2 \left(\Delta_{la}(k_1)\hat{\eta}_{mb}(k_2)+\hat{\eta}_{la}(k_1)\Delta_{mb}(k_2)\right) - \alpha_3\hat{\eta}_{la}(k_1) \hat{\eta}_{mb}(k_2)\right] A^{ab}.
\ea
The previous combination is real and symmetric  under $\vec{k}_1 \leftrightarrow \vec{k}_2$. The function, $A^{ab}$ 
is written as 
\begin{align}
A^{ab}&=A_2(k_1,k_2,k_3)k_2^ak_3^b+A_3(k_1,k_2,k_3)\delta^{ab},
\end{align} 
and is the more general object which does not vanish when contracting with the combination of tensors in the brackets in equation \eqref{z1z1z0de}.
We remark here that another possible term that could be added to the previous solution is $\hat\eta_{ab}(k_3)$, but, it can be checked that this term spoils the symmetry under the change $\vec{k}_1 \leftrightarrow \vec{k}_2$, hence, this term is not  allowed. Following \cite{Bzowski:2013sza}, the solutions for the scalar functions are 
\begin{align}
A_2(k_1,k_2,k_3)&=a_2 J_{2\{000\}},\\
A_3(k_1,k_2,k_3)&=a_2 J_{1\{001\}}+a_3 J_{0\{000\}}, 
\end{align}
then, the second order contribution is
\ba\label{z1z1z0}
&& \langle \zeta^{(1)}(k_1)\zeta^{(1)}(k_2)\zeta^{(0)}(k_3)\rangle +  (\mbox{2 perms.}) = \delta(k_{123})  \times  \\ \nonumber
&& a_2 k_2 k_3 J_{2\{000\}} \left[ (\hat{k}_2\cdot \hat{E}) (\hat{k}_3 \cdot \hat{E})  - (\hat{k}_2\cdot \hat{E})^2 (\hat{k}_2 \cdot \hat{k}_3) -  (\hat{k}_1\cdot \hat{E}) (\hat{k}_3\cdot \hat{E}) (\hat{k}_1 \cdot \hat{k}_2) \right.  \\ \nonumber 
&& \left. +  (\hat{k}_1\cdot \hat{E}) (\hat{k}_2\cdot \hat{E}) (\hat{k}_1 \cdot \hat{k}_2)  (\hat{k}_3 \cdot \hat{k}_2)  - \alpha_3\hat{E}\cdot (\hat{k}_2 \times \hat{k}_1) \hat{E}\cdot (\hat{k}_3 \times \hat{k}_2) \right.  \\ \nonumber
&& \left. i\alpha_2 \left( \hat{E}\cdot (\hat{k}_2 \times \hat{k}_1)(\hat{k}_3 \cdot \hat{E} - (\hat{k}_3 \cdot \hat{k}_2) (\hat{k}_2 \cdot \hat{E}) ) +  \hat{E}\cdot (\hat{k}_3 \times \hat{k}_2)(\hat{k}_2 \cdot \hat{E} - (\hat{k}_1 \cdot \hat{k}_2) (\hat{k}_1 \cdot \hat{E}) ) \right)  \right]  \\ \nonumber
&& + (a_2 J_{1\{001\}}+a_3 J_{0\{000\}}) \left[ 1- (\hat{k}_1 \cdot \hat{E})^2 - (\hat{k}_2 \cdot \hat{E})^2 + (\hat{k}_1 \cdot \hat{k}_2) (\hat{k}_1 \cdot \hat{E}) (\hat{k}_2 \cdot \hat{E}) -\alpha_3 (\hat{E} \times \hat{k}_1)\cdot(\hat{E} \times \hat{k}_2)\right. \\ \nonumber 
&&\left.    -   i \alpha_2 \left(  (\hat{E}\cdot \hat{k}_2) \hat{E} \cdot (\hat{k}_2 \times \hat{k}_1) -  \hat{E} \cdot (\hat{E} \times \hat{k}_1) +  (\hat{E}\cdot \hat{k}_1) \hat{E} \cdot (\hat{k}_1 \times \hat{k}_2) -  \hat{E} \cdot (\hat{E} \times \hat{k}_2)  \right) \right] 
+  (\mbox{2 perms.}).
\ea
In section (\ref{BSST}) we will study the soft limit of this result and propose a general parametrization for the squeezed scalar bispectrum consistent with conformal symmetry. 
\subsection{Tensor-scalar-scalar correlator  }
To finish this section, we study the mixed tensor-scalar-scalar correlator $\langle \gamma \zeta \zeta\rangle$. Mixed correlations such as $\langle \gamma \zeta \zeta\rangle$ have a potential observational interest  and  constitute an important tool to constrain and discriminate inflationary models featuring parity breaking signatures, such as the pseudoscalar coupling $f(\phi)F \tilde{F}$ considered here. In those models the gravitational waves  sourced by the pseudoscalar coupling are chiral  and their presence could leave some distinctive imprints on the correlation functions.  \\ 

As in the previous sections, we will consider the vacuum contribution 
and the leading order contributions coming from the vector field. Therefore, this particular correlator is written as 
\ba
\langle \gamma \zeta \zeta \rangle &=& \langle \gamma^{(0)} \zeta^{(0)}  \zeta^{(0)}  \rangle + \langle \gamma^{(0)} \zeta^{(0)}  \zeta^{(1)} \rangle + \langle \gamma^{(0)} \zeta^{(1)}  \zeta^{(1)} \rangle + \langle \gamma^{(1)}  \zeta^{(1)} \zeta^{(0)} \rangle  + (\mbox{ perms.}).  
\ea
We begin our computation by proposing a general function for the vacuum part, which is consistent with  zero divergence, zero trace and symmetric in the indices $(ij)$.
By using the objects $T^{(n)}_{ijab}$, we propose the following form for the vacuum bispectrum 
\be
\langle \gamma_{ij}^{(0)}(\vec{k}_1) \zeta^{(0)}(\vec{k}_2)  \zeta^{(0)}(\vec{k}_3) \rangle' = B_0 (k_{1}, k_{2}, k_{3}) k_{2a}k_{2b} \left[  T_{ijab}^{(1)}(\vec{k}_1)+ i \beta_1 T_{ijab}^{(2)}(\vec{k}_1) 
\right] \;,
\ee
where $B (k_{1}, k_{2}, k_{3})$ is symmetric under the change $k_2 \leftrightarrow k_3$ and $\beta_{1}$ is a real constant.  The solution  $B(k_{1}, k_{2}, k_{3})$ to the conformal constraints equations was found in \cite{Bzowski:2013sza} and here we use their result. In terms of the triple-K integrals this function is written as   
\be
B_0(k_{1}, k_{2}, k_{3})=b_0 J_{2\{000\}},
\ee
where $b_0$ is a normalization constant. 
%
Projecting into the helicity basis and using the properties \eqref{Piida}-\eqref{eepipi} we obtain 
\be\label{pig0z0z0}
\langle \gamma_{\lambda}^{(0)}(\vec{k}_1) \zeta^{(0)}(\vec{k}_2)  \zeta^{(0)}(\vec{k}_3) \rangle'=\ B_\lambda \Pi_\lambda^{ab}(\vec{k}_1)k_{2a}k_{2b}J_{2\{000\}}\;, \quad {\rm where} \quad B_\lambda = b_0\left[1 + \lambda \beta_1 \right].
\ee
As in the cases involving tensor modes studied previously, this result depends on the helicity and exhibits a different value depending on the polarization.  \\
Now we want to compute the contributions to the correlator sourced by the vector fields up to second order. The first order perturbation $\langle \gamma^{(0)} \zeta^{(1)}  \zeta^{(0)} \rangle$ is related to the correlator $\langle T_{ij} J_{k}  {\cal O} \rangle$ which was calculated in the appendix F of Ref.  \cite{Bzowski:2013sza} and it was determined that it is null for $d \geq 3$\footnote{This conclusion don't change in presence of parity breaking terms since the objects used for the tensorial decomposition in this case obey the same differential Ward identities. See details in Appendix \ref{AI}. }. The leading contributions coming from the vector fields are $\langle \gamma^{(0)} \zeta^{(1)}  \zeta^{(1)} \rangle$  and $\langle \gamma^{(1)} \zeta^{(1)}  \zeta^{(0)} \rangle$.   First, we calculate the term
\be
\langle \gamma_{ij}^{(0)}(\vec{k}_1) \zeta^{(1)}(\vec{k}_2)  \zeta^{(1)}(\vec{k}_3) \rangle\;.
\ee
To do it, we propose a general form consistent with the appropriate symmetry conditions. In this case, we have zero trace, zero divergence and symmetry in the indices $(ij)$ for the momentum $\vec{k}_1$. Additionally, we also have zero divergence in the indices $(lm)$ associated with the vector perturbations in the momentum $\vec{k}_2$ and $\vec{k}_3$. The most general object which is real and fulfill the previous assumptions  has the following form  
\begin{multline} \label{gzaza1}
\langle \gamma_{ij}^{(0)}(\vec{k}_1) \zeta^{(1)}(\vec{k}_2)  \zeta^{(1)}(\vec{k}_3) \rangle=\delta(\vec{k}_{123})\hat{E}^l\hat{E}^m \times \\
\sum_{n=1}^2 \left[ \Delta_{lc}(k_2)\Delta_{md}(k_3)+i \beta_2 (\Delta_{lc}(k_2)\hat{\eta}_{md}(k_3)+\hat{\eta}_{lc}(k_2)\Delta_{md}(k_3))+ \beta_3\hat{\eta}_{lc}(k_2)\hat{\eta}_{md}(k_3)\right] T_{ijab}^{(n)}(\vec{k}_1){\cal B}_{(n)}^{abcd}\;,
\end{multline}
where  ${\cal B}_{(n)}^{abcd}$ are functions which are not annihilated upon contractions with the tensors  $\Delta_{ab}$, $\hat{\eta}_{lc}$ and $T_{ijab}^{(n)}$ in the previous decomposition. Those functions can be written in the form
\begin{equation}\label{snv}
{\cal B}_{(n )}^{abcd} (k_1, k_2, k_3)=B_1^{(n)} k_2^ak_2^bk_3^ck_1^d+B_2^{(n)}\delta^{cd} k_2^ak_2^b+B_3^{(n)}\delta_{ac}k_2^bk_1^d+B_3^{(n)}(k_2 \leftrightarrow k_3)
\delta_{ad}k_2^b k_3^c+B_4^{(n)}\delta_{ac}\delta_{bd},
\end{equation}
where $B_a^{(n)} = B_a^{(n)} (k_1, k_2, k_3)$ and $B_3^{(n)}$ is symmetrized  in $k_2$ and $k_3$. 
It is important to mention here that, differently from the case studied in Ref. \cite{Bzowski:2013sza}, we also need to introduce the odd parity projectors constructed with $\hat\eta_{ab}$ and $T^{(2)}_{abcd}$ and it is not trivial to see that those projectors obey the same equations for the conformal symmetry constraints than the even parity projector $T^{(1)}_{abcd}$. They indeed satisfy the same differential equations as shown in a detailed form in Appendices \ref{AI} and \ref{fm}. Consequently, all the functions ${\cal B}_{(n )}^{abcd}$ also obey the same differential equations, so, we can infer that they are all equal up to relative constants that we will define in the form
\be
{\cal B}_{( 2 )}^{abcd} = \beta_4 {\cal B}_{(1)}^{abcd}, \quad \mbox{wich implies} \quad  B_a^{(2)} = \beta_4 B_a^{(1)}. 
\ee
The solution for $B_a^{(1)}$ was also obtained in Ref. \cite{Bzowski:2013sza}:  
\begin{align}
B_1^{(1)}&= b_1 J_{4\{000\}}\;,\\
B_2^{(1)}&=b_{1} J_{3\{100\}}+b_{2} J_{2\{000\}}\;,\\
B_3^{(1)}&=2b_{1} J_{3\{001\}}+b_{3} J_{2\{000\}}\;,\\
B_4^{(1)}&=2b_{1} J_{2\{011\}}+b_{3} \left(J_{1\{010\}}+J_{1\{001\}}\right)+b_{4} J_{0\{000\}},
\end{align}
where $b_{1,2,3,4}$ are constants. With the previous results, and using the properties \eqref{pt123} we contract with the helicity projector $\Pi_\lambda^{ij}(\vec{k}_1)$ obtaining  
\ba\label{pig0z1z1}
&& \langle \gamma_{\lambda}^{(0)}(\vec{k}_1) \zeta^{(1)}(\vec{k}_2)  \zeta^{(1)}(\vec{k}_3) \rangle' = (1 + \lambda \beta_4 ) \hat{E}^l\hat{E}^m \Pi_\lambda^{ab}(\vec{k}_1) {\cal B}_{(1)}^{abcd}  \times \\ \nonumber
&& \qquad \qquad \qquad  \qquad \qquad \left[ \Delta_{lc}(k_2)\Delta_{md}(k_3)+i \beta_2 (\Delta_{lc}(k_2)\hat{\eta}_{md}(k_3)+\hat{\eta}_{lc}(k_2)\Delta_{md}(k_3))+ \beta_3\hat{\eta}_{lc}(k_2)\hat{\eta}_{md}(k_3)\right] . 
\ea\\
Finally, we compute the contribution from 
\be
\langle \gamma_{ij}^{(1)}(\vec{k}_1) \zeta^{(1)}(\vec{k}_2)  \zeta^{(0)}(\vec{k}_3) \rangle\;,
\ee
and,  proceeding in the same form as we did before, we construct the most general real object consistent with the divergence and trace free conditions and with the symmetries in the  appropriate combination of indices. For this case we have:
\begin{multline} \label{gz1z1z0}
\langle \gamma_{ij}^{(1)}(\vec{k}_1) \zeta^{(1)}(\vec{k}_2)  \zeta^{(0)}(\vec{k}_3) \rangle=\delta(\vec{k}_{123})\hat{E}^l\hat{E}^m \times \\
\sum_{n=1}^2 \left[ \Delta_{lc}(k_1)\Delta_{md}(k_2)+i \beta_5 (\Delta_{lc}(k_1)\hat{\eta}_{md}(k_2)+\hat{\eta}_{lc}(k_1)\Delta_{md}(k_2))+ \beta_6\hat{\eta}_{lc}(k_1)\hat{\eta}_{md}(k_2)\right] T_{ijab}^{(n)}(\vec{k}_1) \bar{{\cal B}}_{(n)}^{abcd}\;,
\end{multline}
where
\begin{equation}\label{snv1}
\bar{{\cal B}}_{(n)}^{abcd} =\bar{B}_1^{(n)} k_2^ak_2^bk_2^ck_3^d+ \bar{B}_2^{(n)}\delta^{cd} k_2^ak_2^b+ \bar{B}_3^{(n)}\delta^{ac}k_2^bk_3^d+ \bar{B}_4^{(n)}\delta^{ad}k_2^b k_2^c+ \bar{B}_5^{(n)}\delta^{ac}\delta^{bd},
\end{equation}
and $\bar{B}_1^{(n)} = \bar{B}_1^{(n)}(k_1, k_2, k_3)$ are functions of the momenta. As before, given that all the functions $\bar{{\cal B}}_{(n )}^{abcd}$ obey the same differential equations, they are all equal up to relative constants that we will define in the form
\be
\bar{{\cal B}}_{( 2 )}^{abcd} = \beta_7 \bar{{\cal B}}_{(1)}^{abcd}, \quad \mbox{which implies} \quad \bar{B}_a^{(2)} = \beta_7 \bar{B}_a^{(1)}. 
\ee
Again, we follow the techniques described in \cite{Bzowski:2013sza} to find the solutions for the functions $\bar{B}_a^{(1)}(k_1, k_2, k_3)$ obtaining:    
\begin{align}
\bar{B}_1^{(1)}&= b_5 J_{4\{000\}}\;,\\
\bar{B}_2^{(1)}&= \bar{B}_4^{(1)} = b_{6} J_{2\{000\}}\, ,\\ 
\bar{B}_3^{(1)}&=2b_{5} J_{3\{010\}}+b_{7} J_{2\{000\}}\;,\\
\bar{B}_5^{(1)}&=b_{6} J_{1\{010\}}+b_{8} J_{0\{000\}},
\end{align} This particular case is not reported in Ref. \cite{Bzowski:2013sza}. The explicit and detailed form in which we derived the previous solution can be found in Appendix  \ref{fm}. Finally, we project with the operator $\Pi_\lambda^{ij}(\vec{k}_1)$  and use the properties  \eqref{pt123} to obtain
\ba \label{pig1z1z0}
&& \langle \gamma_{\lambda}^{(1)}(\vec{k}_1) \zeta^{(1)}(\vec{k}_2)  \zeta^{(0)}(\vec{k}_3) \rangle'=(1+ \lambda \beta_7 ) \hat{E}^l\hat{E}^m \Pi_\lambda^{ab}(\vec{k}_1) \bar{{\cal B}}_{(1)}^{abcd}  \times \\ \nonumber
&&  \qquad \qquad \qquad \qquad \qquad  \left[ \Delta_{lc}(k_1)\Delta_{md}(k_2)+i \beta_5 (\Delta_{lc}(k_1)\hat{\eta}_{md}(k_2)+\hat{\eta}_{lc}(k_1)\Delta_{md}(k_2))+ \beta_6\hat{\eta}_{lc}(k_1)\hat{\eta}_{md}(k_2)\right].
\ea
Next section is devoted to study the soft limit of the correlators considered in this section. 
\section{The squeezed configuration and consistency relations }
\label{BSST}
It is instructive to evaluate the soft limits of the previous results. Soft limits, in which one or a combination of momenta go to zero, are relevant to constrain the particle content during inflation \cite{Arkani-Hamed:2015bza}. In particular, a significant (measurable) value of the squeezed limit of the three point function can be seen as a signature of the presence of light and long lived degrees of freedom during the inflationary expansion \cite{Maldacena:2002vr,Creminelli:2004yq,Cheung:2007sv,Li:2008gg,Assassi:2012zq,Creminelli:2012ed,Anninos:2019nib}. We evaluate the squeezed limit and derive consistency relations for the particular model which we have discussed here. 
\subsection{The scalar bispectrum $\langle \zeta \zeta \zeta \rangle$ }
To start, let's evaluate the squeezed limit of the scalar bispectrum. For the vacuum contribution of the scalar bispectrum we take the limit as $k_1\rightarrow 0$ of the result \eqref{z0z0z0}. 
In that limit, $k_1\rightarrow 0 $ and $k_2\rightarrow k_3$ and we obtain
\ba\nonumber \label{z0z0z01}
\lim_{k_1 \rightarrow 0} \langle \zeta^{(0)}(k_1)\zeta^{(0)}(k_2)\zeta^{(0)}(k_3)\rangle' &\propto& a_0 k_1^{-3+2\Delta_{\zeta0}} k_2^{-3+\Delta_{\zeta0}} \\ 
&\propto& a_0  k_2^{-\Delta_{\zeta0}}  \langle \zeta^{(0)}(k_1)\zeta^{(0)}(-k_1)\rangle \langle \zeta^{(0)}(k_2)\zeta^{(0)}(-k_2)\rangle. 
\ea
Now, we calculate the squeezed limit $\vec{k}_1 \rightarrow 0$ of the sourced contributions to the scalar bispectrum. In this limit, the contribution from $J_{2\{0 0 0\}}$ and $J_{1\{0 0 0\}}$ vanishes, so, the only contribution comes from $J_{0\{0 0 0\}}$ and then 
\be
B_2(k_1,k_2,k_3) \approx a_3 k_1^{-3+2\Delta_{\zeta0}+2\Delta_{u}}k_2^{-3+\Delta_{\zeta0}}.
\ee
In this case, the scalar bispectrum can be written as
\begin{multline}\label{zazaz}
\lim_{k_1 \rightarrow 0}\langle \zeta^{(1)}(k_1)\zeta^{(1)}(k_2)\zeta^{(0)}(k_3)\rangle' + (\mbox{2 perms.}) \propto   k_1^{-3+2\Delta_{\zeta0}+2\Delta_{u}}k_2^{-3+\Delta_{\zeta0}}\times \\
\left[ 2\left(1-(\hat{E}\cdot \hat{k}_1)^2-(\hat{E}\cdot \hat{k}_2)^2+(\hat{k}_2\cdot \hat{k}_1)(\hat{E}\cdot \hat{k}_1)(\hat{E}\cdot \hat{k}_2)\right) - 2 \alpha_3\left(\hat{k}_1\cdot\hat{k}_2-(\hat{E}\cdot \hat{k}_1)(\hat{E}\cdot \hat{k}_2)\right)\right. \\
\left. -2 i\alpha_2 \left(\hat{E}\cdot(\hat{k}_1-\hat{k}_2)\hat{E}\cdot(\hat{k}_1 \times \hat{k}_2)\right) + \left( 1 - (\hat{E}\cdot \hat{k}_2)^2 +\alpha_3 (\hat{E}\times\hat{k}_2)^2\right) \right].
\end{multline}
Averaging over the vector $\hat{E}$, that is, integrating over all the possible directions of this vector $\int \frac{d^2 \Omega}{4\pi}$ we obtain
\begin{multline} \label{averagez0z1z1}
 \lim_{k_1 \rightarrow 0}\langle \zeta^{(1)}(k_1)\zeta^{(1)}(k_2)\zeta^{(0)}(k_3)\rangle' + (\mbox{2 perms.}) \\ 
 \propto   k_1^{-3+2\Delta_{\zeta0}+2\Delta_{u}}k_2^{-3+\Delta_{\zeta0}}  \left[1+\frac{\alpha_3}{2} -  \alpha_3 \hat{k}_1\cdot \hat{k}_2  +\frac{1}{2} (\hat{k}_1\cdot \hat{k}_2)^2  \right] \\ 
 \propto  k_2^{-\Delta_{\zeta0}} k_1^{2\Delta_{u}}  \langle \zeta^{(0)}(k_1)\zeta^{(0)}(-k_1)\rangle \langle \zeta^{(0)}(k_2)\zeta^{(0)}(-k_2)\rangle \times\\ 
 \left[ \left(\frac{7 + 3 \alpha_3 }{6}  \right) P_{0}(\hat{k}_1\cdot \hat{k}_2) - \alpha_3 P_{1}(\hat{k}_1\cdot \hat{k}_2) +  \frac{1}{3} P_{2}(\hat{k}_1\cdot \hat{k}_2) \right],
\end{multline}
which depends on the magnitude of $\vec{k}_1$ and $\vec{k}_2$ and also on the angle between them  $(\hat{k}_1\cdot \hat{k}_2) = \cos \hat{\theta}$. In the last line of the previous expression we wrote the polynomial of $\cos \hat{\theta}$ in terms of the Legendre polynomials $P_{L}$ with $L=0,1,2$.  Finally, it is easy to see that the angle average of the contributions from $\langle \zeta^{(1)} \zeta^{(0)} \zeta^{(0)}\rangle$ in \eqref{z1z0z0}, is zero. With this, adding the vacuum result \eqref{z0z0z01} with the sourced contributions in \eqref{averagez0z1z1} we obtain  
\ba\label{limzzz}
\lim_{k_1 \rightarrow 0} \langle \zeta (\vec{k}_1)  \zeta (\vec{k}_2)  \zeta (\vec{k}_3)  \rangle &\propto& k_2^{-\Delta_{\zeta0}} P_{\zeta}(k_1) P_{\zeta}(k_2) \sum_{l=0}^2 c_L (k_1, k_2) P_{L} (\hat{k}_1\cdot \hat{k}_2)
\ea
where
\be
c_{0} = 1+ \frac{a_3}{a_0} \left( \frac{7 + 3\alpha_3}{6} \right)  k_{1}^{2\Delta_{u}}, \qquad c_{1} = -\alpha_3 \frac{a_3}{a_0}k_{1}^{2\Delta_{u}}, \qquad c_{2} = \frac{a_3}{3a_0}k_{1}^{2\Delta_{u}}.
\ee
It is interesting to notice that the angular dependence introduced here is a direct consequence of the presence of a vector field. The precise form of the angular dependence of the three point function provides relevant information about the presence of vector fields and about their interactions with other fields. Additionally, the coefficients $c_{L} (k_1, k_2)$ acquire scale dependence through the scaling dimension of the vector perturbations in the form $c_{L} \sim k_{1}^{2\Delta_{u}} \sim k_{1}^{4+2n}$. This dependence exhibits scale invariant behaviour at $n=-2$. Nevertheless, when the power of the interaction term \eqref{coup} goes like $n\gtrsim -2$, some level of scale dependence is obtained which can be seen as distinctive feature of the model.   The expression \eqref{limzzz} constitutes a useful expression to parametrize departures from conformal symmetry in scenarios allowing for general interactions between scalar and vector fields through the coupling term \eqref{coup}. It is interesting to notice that  the term $P_{1}(\hat{k}_1\cdot \hat{k}_2)$ is not symmetric under the reflection $\hat{\theta} \rightarrow \pi -\hat{\theta} $ which is a consequence of the presence of the parity breaking  term $f(\phi) \tilde{F}F$.  A complete calculation, including the amplitude factors in the case with $n=-2$, was performed in \cite{Bartolo:2015dga} by using the in-in formalism. Our results here, agree with the results in \cite{Bartolo:2015dga} up to amplitude factors in the case $n=-2$. 

\subsection{The mixed bispectrum $\langle \gamma \zeta \zeta \rangle$ }
To start, we evaluate the squeezed limit of the vacuum contribution \eqref{pig0z0z0}  of the mixed correlator.  
Taking the limit $\vec{k}_1 \rightarrow 0$ and evaluating the integral, \eqref{pig0z0z0} becomes
\ba\label{limg0z0z0}
 \lim_{k_1 \rightarrow 0}\langle \gamma_{\lambda}^{(0)}(\vec{k}_1) \zeta^{(0)}(\vec{k}_2)  \zeta^{(0)}(\vec{k}_3) \rangle' &\propto& B_{0\lambda}  k_1^{-3+2\Delta_{\gamma0}}k_2^{-3-\Delta_{\gamma0}+2\Delta_{\zeta0} }  \Pi_\lambda^{ab}(\vec{k}_1)\hat{k}_{2a}\hat{k}_{2b}  \\ \nonumber
&\propto&  B_{0\lambda} k_2^{-\Delta_{\gamma0}}   \langle \gamma_{\lambda}^{(0)}(\vec{k}_1) \gamma_{\lambda}^{(0)}(-\vec{k}_1) \rangle \langle \zeta^{(0)}(\vec{k}_2) \zeta^{(0)}(-\vec{k}_2) \rangle  \Pi_\lambda^{ab}(\vec{k}_1)\hat{k}_{2a}\hat{k}_{2b},
\ea 
which is a very well known result that we reproduce here up to normalization factors, see e.g. \cite{Maldacena:2002vr} and more recently 
in  \cite{Anninos:2019nib}  in the context of inflationary models that include 
higher spin fields.
Now, we add the contributions from the sourced parts $\langle \gamma_{\lambda}^{(0)} \zeta^{(1)}  \zeta^{(1)}  \rangle$ and $\langle \gamma_{\lambda}^{(1)} \zeta^{(1)}  \zeta^{(0)}  \rangle$. To begin, we study the limit $\vec{k}_1 \rightarrow 0$ and $\vec{k}_3 \rightarrow -\vec{k}_2$ of \eqref{pig0z1z1}. First, we notice that in this limit the contribution from $B_1^{(1)}$ is accompanied by $k_2^ak_2^bk_2^ck_1^d$ which vanishes when contracted with the projectors $\Delta_{lc}(k_2)$ and $\hat{\eta}_{ac}(k_2)$. So, we only need to consider the contributions from $B_2^{(1)},  B_3^{(1)}, B_4^{(1)}$. Using the approximation of the triple-K integrals in the limit $k_1\rightarrow 0$ we obtain
\ba
B_2^{(1)}&\approx&
 \sigma_{2}   k_1^{-3+2\Delta_{\gamma 0}} k_2^{-5+2 \Delta_{\zeta0} -\Delta_{\gamma 0} + 2\Delta_{u}}, \\  
B_3^{(1)}&\approx& \sigma_{3}   k_1^{-3+2\Delta_{\gamma 0}} k_2^{-5+2 \Delta_{\zeta0} -\Delta_{\gamma 0} + 2\Delta_{u}} 
\\
B_4^{(1)}&\approx& \sigma_{4}  k_1^{-3+2\Delta_{\gamma 0}} k_2^{-3+2 \Delta_{\zeta0} -\Delta_{\gamma 0} + 2\Delta_{u} }, 
\ea 
where we have redefined the constants $b_{1,2,3,4}$ in terms of the constants $\sigma_{2,3,4}$ which carry some dependence of the conformal weights through the gamma functions appearing in the expansions of the Bessel K functions. By using the previous results and neglecting the terms with $(k_1/k_2)$ and $(k_1/k_2)^2$, 
we find that
\be
{\cal B}_{(1)}^{abcd} \approx  \sigma_{2}   k_1^{-3+2\Delta_{\gamma 0}} k_2^{-3+2 \Delta_{\zeta0} -\Delta_{\gamma 0} + 2\Delta_{u}} \left( \delta^{cd} \hat{k}_{2}^{a} \hat{k}_{2}^{b}  + \frac{ \sigma_{4} }{ \sigma_{2} } \delta^{ac} \delta^{bd} \right).
\ee
Inserting it into \eqref{pig0z1z1} we obtain 
\ba
&& \lim_{k_1 \rightarrow 0} \langle \gamma_{\lambda}^{(0)}(\vec{k}_1) \zeta^{(1)}(\vec{k}_2)  \zeta^{(1)}(\vec{k}_3) \rangle' = (1 + \lambda \beta_4 )\sigma_{2}   k_1^{-3+2\Delta_{\gamma 0}} k_2^{-3+2 \Delta_{\zeta0} -\Delta_{\gamma 0} + 2\Delta_{u}}   \times \\ \nonumber
&& \Pi_\lambda^{ab}(\vec{k}_1) \hat{k}_{2a} \hat{k}_{2b}  (1- \beta_3) (1- (\hat{E}\cdot \hat{k}_2)^2) \\ \nonumber 
&& +  \frac{ \sigma_{4} }{ \sigma_{2} }   \Pi_\lambda^{ab}(\vec{k}_1)   \left[ (\hat{E}_{a} - (\hat{E}\cdot \hat{k}_2)\hat{k}_{2a})(\hat{E}_{b} - (\hat{E}\cdot \hat{k}_2)\hat{k}_{2b}) + \beta_4  (\hat{E}\times \hat{k}_{2})_{a}(\hat{E}\times \hat{k}_{2})_{b} \right] \\ \nonumber
&& \propto k_2^{-\Delta_{\gamma0}+ 2\Delta_{u} }   \langle \gamma_{\lambda}^{(0)}(\vec{k}_1) \gamma_{\lambda}^{(0)}(-\vec{k}_1) \rangle \langle \zeta^{(0)}(\vec{k}_2) \zeta^{(0)}(-\vec{k}_2) \rangle  \Pi_\lambda^{ab}(\vec{k}_1)\times \\ \nonumber
&&  \left[  \hat{k}_{2a}\hat{k}_{2b}(1- (\hat{E}\cdot \hat{k}_2)^2) + \sigma  \left[ (\hat{E}_{a} - (\hat{E}\cdot \hat{k}_2)\hat{k}_{2a})(\hat{E}_{b} - (\hat{E}\cdot \hat{k}_2)\hat{k}_{2b}) + \beta_4  (\hat{E}\times \hat{k}_{2})_{a}(\hat{E}\times \hat{k}_{2})_{b} \right] \right]
\ea
where we  defined the constant $\sigma = \sigma_{4}/(\sigma_{2}(1- \beta_3))$. 
Finally, integrating over all the possible angles described by the vector $\hat{E}$ we obtain 
\ba\label{limg1z0z0}
\lim_{k_1 \rightarrow 0} \langle \gamma_{\lambda}^{(0)}(\vec{k}_1) \zeta^{(1)}(\vec{k}_2)  \zeta^{(1)}(\vec{k}_3) \rangle' &\propto& \\ \nonumber
&&  B_{1\lambda} k_2^{-\Delta_{\gamma0}+ 2\Delta_{u} }   \langle \gamma_{\lambda}^{(0)}(\vec{k}_1) \gamma_{\lambda}^{(0)}(-\vec{k}_1) \rangle \langle \zeta^{(0)}(\vec{k}_2) \zeta^{(0)}(-\vec{k}_2) \rangle    \Pi_\lambda^{ab}(\vec{k}_1) \hat{k}_{2a} \hat{k}_{2b}. 
\ea
Now, following the same procedure , we take the squeezed limit of \eqref{pig1z1z0}. First, we integrate each of the non-vanishing terms in Eq. (\ref{snv1}), which gives
\ba
\bar{B}_2^{(1)}&\approx&\bar{\sigma}_{2} k_2^{-5-\Delta_{\gamma0}+2\Delta_{\zeta0} }k_1^{-3+2\Delta_{\gamma0} + 2 \Delta_{u}} \\
\bar{B}_5^{(1)} &\approx& \bar{\sigma}_{5}k_2^{-3-\Delta_{\gamma0}+2\Delta_{\zeta0} }k_1^{-3+2\Delta_{\gamma0} + 2 \Delta_{u}},
\ea
where $\bar{\sigma}_{2,5}$ are constants defined in terms of the constants ${b}_{6,8}$ and the gamma functions which appear in the expansions of the Bessel K functions.  Using these results, the Eq. (\ref{snv1})  is written as
\begin{equation}\label{snvsq1}
\bar{{\cal B}}_{(1)}^{abcd} \approx \bar{\sigma}_{2}  k_2^{-3-\Delta_{\gamma0}+2\Delta_{\zeta0} }k_1^{-3+2\Delta_{\gamma0} + 2 \Delta_{u}} \left[\delta^{cd} \hat{k}_2^a\hat{k}_2^b  +\delta^{ad}\hat{k}_2^b\hat{k}_2^c  +\frac{\bar{\sigma}_{5}}{ \bar{\sigma}_{2}}\delta^{ac}\delta^{bd}  \right],
\end{equation}
and inserting this expression into \eqref{pig1z1z0} we obtain
 \begin{multline} \label{gzaza6}
 \lim_{k_1 \rightarrow 0}  \langle \gamma_{\lambda}^{(1)}(\vec{k}_1) \zeta^{(1)}(\vec{k}_2)  \zeta^{(0)}(\vec{k}_3) \rangle' \propto B_{2\lambda} k_2^{-3-\Delta_{\gamma0}+2\Delta_{\zeta0} }k_1^{-3+2\Delta_{\gamma0} + 2 \Delta_{u}}  \left \{ \right. \\
 \Pi_{ab}^{\lambda}(\vec{k}_1)\hat{k}_2^a\hat{k}_2^b\left[1-(\hat{E}\cdot \hat{k}_1)^2-2(\hat{E}\cdot \hat{k}_2)^2+2(\hat{E}\cdot \hat{k}_1)(\hat{E}\cdot \hat{k}_2)(\hat{k}_1\cdot \hat{k}_2)+\beta_6\left( (\hat{k}_1\cdot \hat{k}_2)-(\hat{E}\cdot \hat{k}_2)(\hat{E}\cdot \hat{k}_1) \right) \right.\\
\left.+i \beta_{5} \hat{E}\cdot(2\hat{k}_2-\hat{k}_1)\hat{E}\cdot (\hat{k}_1 \times \hat{k}_2) \right]+\\
 \Pi_{ab}^{\lambda}(\vec{k}_1)\hat{k}_2^b \hat{E}^{a}  \left((\hat{E}\cdot \hat{k}_2)-(\hat{E}\cdot \hat{k}_1)(\hat{k}_1\cdot \hat{k}_2)-i \beta_5 \hat{E}\cdot (\hat{k}_1\times \hat{k}_2)\right)+\\
 \Pi_{ab}^{\lambda}(\vec{k}_1)\hat{k}_2^b (\hat{E}\times k_{2})^{a}  \left( \beta_6\hat{E}\cdot (\hat{k}_1\times \hat{k}_2 ) -i\beta_5 (\hat{E}\cdot \hat{k}_2 - (\hat{k}_1\cdot \hat{k}_2 )(\hat{E}\cdot \hat{k}_1))\right)+\\
\bar{\sigma}   \Pi_{ab}^{\lambda}(\vec{k}_1)\left[ \hat{E}^a \hat{E}^b - \hat{E}^a  \hat{k}_2^{b} (\hat{E}\cdot \hat{k}_2)  + \beta_6 (\hat{E} \times \hat{k}_1)^{a} (\hat{E} \times \hat{k}_2)^{b} \right. \\
\left. -i\beta_5 \left( \hat{E}^{a} (\hat{E} \times \hat{k}_2)^{b}  + \hat{E}^{b} (\hat{E} \times \hat{k}_1)^{a} - \hat{k}_{2}^{b} (\hat{E} \times \hat{k}_1)^{a} (\hat{E}\cdot \hat{k}_{2}) \right)  \right] \left. \right \}.
\end{multline}
where $\bar{\sigma} = \bar{\sigma}_{5}/\bar{\sigma}_{2}$. Integrating over all the possible directions of the vector $\hat{E}$ we obtain
 \ba\label{limg1z0z0intE}\nonumber
 \lim_{k_1 \rightarrow 0}  \langle \gamma_{\lambda}^{(1)}(\vec{k}_1) \zeta^{(1)}(\vec{k}_2)  \zeta^{(0)}(\vec{k}_3) \rangle' &\propto& B_{2\lambda} k_2^{-\Delta_{\gamma0}}  k_1^{2\Delta_{u}}    \langle \gamma_{\lambda}^{(0)}(\vec{k}_1) \gamma_{\lambda}^{(0)}(-\vec{k}_1) \rangle \langle \zeta^{(0)}(\vec{k}_2) \zeta^{(0)}(-\vec{k}_2) \rangle \Pi_{ab}^{\lambda}(\vec{k}_1)\hat{k}_2^a\hat{k}_2^b \times   \\ 
&&  \frac{1}{3} \left[ 1- \bar{\sigma} + 2(\hat{k}_1\cdot \hat{k}_2 )^2 + 3 \beta_6  (\hat{k}_1\cdot \hat{k}_2) + \lambda \beta_5 \left(1-  (\hat{k}_1\cdot \hat{k}_2) + \bar{\sigma}\right)\right] .   %
\ea
Finally, we sum up the contributions from \eqref{limg0z0z0}, \eqref{limg1z0z0} and \eqref{limg1z0z0intE}  to obtain the squeezed limit of the full correlator: 
 \ba\label{limgzzintE}
 \lim_{k_1 \rightarrow 0}  \langle \gamma_{\lambda}(\vec{k}_1) \zeta (\vec{k}_2)  \zeta(\vec{k}_3) \rangle' &\propto&  B_{0\lambda} k_2^{-\Delta_{\gamma0}}      \langle \gamma_{\lambda}^{(0)}(\vec{k}_1) \gamma_{\lambda}^{(0)}(-\vec{k}_1) \rangle \langle \zeta^{(0)}(\vec{k}_2) \zeta^{(0)}(-\vec{k}_2) \rangle \Pi_{ab}^{\lambda}(\vec{k}_1)\hat{k}_2^a\hat{k}_2^b \times   \\  \nonumber
 &&    \frac{1}{3} \left[1+ V_{1\lambda}k_2^{2\Delta_{u}}  +k_1^{2\Delta_{u}} V_{2\lambda} \left[  \lambda \beta_5 \left( (1+\bar{\sigma})P_0(\hat{k}_1\cdot \hat{k}_2 ) - P_1(\hat{k}_1\cdot \hat{k}_2 ) \right)\right. \right. + \\ \nonumber
 && \left. \left.  \left( \left(\frac{5}{3} - \bar{\sigma}\right) P_{0}(\hat{k}_1\cdot \hat{k}_2 ) + 3 \beta_6 P_{1}(\hat{k}_1\cdot \hat{k}_2)  + \frac{4}{3}P_2(\hat{k}_1\cdot \hat{k}_2 ) \right) \right] \right],
\ea
where we have expressed the polynomial in the angle $\hat{k}_1\cdot \hat{k}_2$ in terms of the Legendre polynomials $P_{L=0,1,2}$. In a generic form, we can write 
 \ba\label{limgzzintELP}
 \lim_{k_1 \rightarrow 0}  \langle \gamma_{\lambda}(\vec{k}_1) \zeta (\vec{k}_2)  \zeta(\vec{k}_3) \rangle' &\propto&  B_{0\lambda} k_2^{-\Delta_{\gamma0}}      \langle \gamma_{\lambda}^{(0)}(\vec{k}_1) \gamma_{\lambda}^{(0)}(-\vec{k}_1) \rangle \langle \zeta^{(0)}(\vec{k}_2) \zeta^{(0)}(-\vec{k}_2) \rangle \Pi_{ab}^{\lambda}(\vec{k}_1)\hat{k}_2^a\hat{k}_2^b \times   \\  \nonumber
 &&  \sum_{L=0}^2 c^{\lambda}_{L}(k_1, k_2) P_{L}(\hat{k}_1\cdot \hat{k}_2 ).
\ea
The previous expression for the parametrization of the tensor-scalar-scalar correlator in the squeezed limit has some interesting features that is worth to mention. As  in the case of the scalar bispectrum seen before, the coefficients $c^{\lambda}_{L} (k_1, k_2)$ acquire scale dependence through the scaling dimension of the vector perturbations in the form $c^{\lambda}_{L} \sim k^{4+2n}$.  Additionally, the coefficients $c^{\lambda}_{L} (k_1, k_2)$ also depend on the particular polarization, establishing in this way an explicit difference in the correlations $\langle \gamma_{+}\zeta \zeta \rangle $  and $\langle \gamma_{-}\zeta \zeta \rangle$. This implies that the angular dependence and the expansion in Legendre polynomials is different for each polarization.  This constitutes another distinctive feature of the model considered here and could be of potential observational interest.  We remark that we derived the previous expression imposing only nearly conformal invariance of the scalar, vector and tensor perturbations.  

\section{Conclusion and final remarks}\label{Conclusions}
In this paper we have explored the consequences of imposing conformal invariance in the inflationary scalar vector model described by the interaction Lagrangian \eqref{coup}. Following the analysis done in \cite{Biagetti:2013qqa} for the model $f(\phi)F^2$ and in \cite{Almeida:2017lrq} for the model $f(\phi)(F^2+ \alpha F\tilde{F})$, here we used CFT methods to constrain the shape of two and three points inflationary correlators involving scalar, vector and tensor perturbations.\\ 

For the analysis of the three point functions we rely on the results reported by  Bzowski {\it et al} in Ref. \cite{Bzowski:2013sza}. This reference offers a complete and detailed list of results for the two and three point functions involving scalar, vector and tensor operators, in any dimension, in momentum space and restricted to parity conserving systems. 
Here we have extrapolated the results of this reference to include the parity breaking case, supported for the model \eqref{coup}, which enjoys super horizon conformal symmetry \cite{Valenzuela2016, Almeida:2017lrq}. One of the main difficulties that we found in applying the results from  \cite{Bzowski:2013sza} to our case, was to obtain a complete tensor decomposition for the three point functions 
in momentum space in such a way that the parity breaking contributions were  properly taken into account. In section \ref{spectrum} we obtain the tensors $T^{(a)}_{ijkl}(\vec{k})$ which respects $O(3)$ symmetry and reflect the gauge choice of being divergence free and trace free in the appropriate combination of indices $(ijkl)$. Those objects include parity breaking contributions by including the Levi-Civita tensor $\eta_{ijk}$ in their structure. All the elements necessary to express the two and three point functions can be written as a combination of the objects  $T^{(a)}_{ijkl}(\vec{k}),  \Delta_{ij}(\vec{k})$ and $\hat{\eta}_{ij}(\vec{k})$. Interestingly enough, we've found that the tensors $T^{(a)}_{ijkl}(\vec{k})$ has the property of obeying the same differential equations that their parity symmetric counterparts and this property allows us to extend the results obtained in  \cite{Bzowski:2013sza} to our case, just  by adding, to the parity preserving decomposition tensor structure, the appropriate combination of $T^{(a)}_{ijkl}(\vec{k}),  \Delta_{ij}(\vec{k})$ and $\hat{\eta}_{ij}(\vec{k})$.   \\
 
With the objects  mentioned before, we construct the two and three point correlators. Our main results can be found in section \ref{bispectrum}. In this section we obtain a general expression for the scalar bispectrum $\langle \zeta \zeta \zeta \rangle $ in equation \eqref{z1z1z0} and for the mixed tensor-scalar-scalar correlator $\langle \gamma \zeta \zeta \rangle $ in equations \eqref{pig0z1z1} and \eqref{pig1z1z0}. The squeezed limit of those expressions, after angle averaging over the VEV direction, offers a useful parametrization which reveals interesting features such as scale dependence as a power law in terms of the conformal weight $\Delta_u$  of the vector perturbations. Aside from this, we derived general expressions for the angular dependence of the three point functions in this limit. The case of the  tensor-scalar-scalar correlator $\langle \gamma \zeta \zeta \rangle $ is 
interesting  because it 
displays 
a particular dependence on the polarization of the tensor mode. In this case, not only the amplitudes but also the angular dependence expressed by the expansion of Legendre polynomials are different for the two polarizations. We consider that those particular features of the correlator  $\langle \gamma \zeta \zeta \rangle $ could be relevant  for 
seeking possible parity breaking patterns in the CMB. The correlators  $\langle   \gamma \gamma \zeta \rangle$  and  $\langle \gamma \gamma \gamma \rangle$ would also reveal important information about the mechanism of sourcing of chiral gravitational waves. Nevertheless, we don't include explicit details of such correlators here since it would involve the construction of tensor structures, similar to the ones listed in Appendix \ref{umatrices}, but with 8 indices. This construction is relevant and interesting but this  goes a bit further from our main purpose with the present work which is to show the use of conformal symmetries  as a guide to reveal aspects of the  structure of the correlators of cosmological perturbations. We consider that this is illustrated with enough detail with the correlators studied in the paper. \\

Finally, as mentioned previously, we should emphasize that the formalism that we followed here does not allow us to find exact and complete expressions for the three point correlators.   Our results are expressed in terms of a set of arbitrary constants that are not determined only by symmetry considerations. Nevertheless, we illustrated here that the use of symmetries  is a fundamental guiding principle that allows us to reveal general characteristics and features of the inflationary correlators. One of the most remarkable results that we found here, is that we were able to find the relevant scale dependence in the squeezed limit in terms of the power $n$ of the coupling function $f(\phi)$. Additionally,  the angular and the helicity dependence of the correlators were obtained  relying only on the symmetries of the system.

\section*{Acknowledgments}
This work was supported by COLCIENCIAS grant  110671250405 RC FP44842-103-2016, COLCIENCIAS-DAAD grant 110278258747 RC-774-2017 and  by Universidad Antonio Nari\~no grant number 2017239. JPBA thanks Universidad del Valle for its warm hospitality during several stages of this project.

\appendix
\section{Tensor decomposition consistent with the symmetries and the gauge choice}\label{umatrices}
In this appendix we show in a broad manner the procedure to build the tensors with four indices $B^{(n)}_{ijlm}$ \eqref{Bijred1}-\eqref{Bijred4} which are invariant under $O(3)$ group and  that reflect the gauge conditions of the perturbations 
$\nabla^{i}\delta {E}_{i} = \nabla^{i} \gamma_{ij} =\nabla^{j} \gamma_{ij} =0$ for the corresponding indices. The tensors $B^{(n)}_{ijlm}$ appear in the construction of the correlator 
\be 
\langle \gamma_{ij}^{(1)} (\vec{k}_1) \zeta^{(1)} (\vec{k}_2) \rangle = \delta(\vec{k}_{12})\hat{E}^l\hat{E}^m{\rm B}_{ijlm}.  
\ee
On the left hand side of this correlator we have the tensor $\gamma_{ij}$ and two vector perturbations $\delta E_{l}$ and $\delta E_{m}$, which, according with \eqref{pertdA} enter in the correlator in the form
\be 
\langle \gamma_{ij}^{(1)} (\vec{k}_1) \zeta^{(1)} (\vec{k}_2) \rangle \sim  \hat{E}^l\hat{E}^m \langle (\delta E_{l} \gamma_{ij}^{(0)} )(\vec{k}_1) (\delta E_{m} \zeta^{(0)} )(\vec{k}_2) \rangle.  
\ee
Then, we need an object with the following properties: divergence free in all the indices $(ijlm)$, symmetric and trace free in $(ij)$.  For the two point correlator, we can think that all the conditions are evaluated for the same momentum $\vec{k}_1$ or $\vec{k}_2$ since momentum conservation implies $\vec{k}_1 = - \vec{k}_2$. For the properties mentioned before we can use the tensors $\Delta_{ij}$ and $\hat{\eta}_{ij}$ as the starting point of our construction since they are zero divergence for all the indices $(ijlm)$. In general we can write the tensor ${\rm B}_{ijlm}$ as an expansion of $\Delta$'s and $\hat{\eta}$'s in the form:
 \begin{align}
{\rm B}_{ijlm} = & b_1 \Delta_{ij}\Delta_{lm}+b_2 \Delta_{il}\Delta_{jm}+b_3 \Delta_{im}\Delta_{lj}  + b_4 \hat{\eta}_{ij}\Delta_{lm}+b_5 \hat{\eta}_{il}\Delta_{jm}+b_6 \hat{\eta}_{im}\Delta_{lj}\nonumber\\
+ &  b_7 \Delta_{ij}\hat{\eta}_{lm}+b_8 \Delta_{il}\hat{\eta}_{jm}+b_9 \Delta_{im}\hat{\eta}_{lj} + b_{10} \hat{\eta}_{ij}\hat{\eta}_{lm}+b_{11} \hat{\eta}_{il}\hat{\eta}_{jm}+b_{12} \hat{\eta}_{im}\hat{\eta}_{lj}.\nonumber
\end{align}
Imposing symmetry in $(ij)$ we obtain $b_2 = b_3, \, b_5 = -b_9, \, b_6=b_8,\, b_{11}=-b_{12}, \, b_4 = 0$ and $b_{10} = 0.$ This reduces the number of independent combinations to six. Furthermore, imposing zero trace in $(ij)$ we obtain $b_{11} = -(b_1 + b_2)$ and $b_{5} = b_6 + b_7$, which reduces the number of independent tensors to four. Finally, using cyclic properties of the Levi-Civita symbol in three dimensions, we can see that 
\be
\hat{\eta}_{il}\hat{\eta}_{jm}+ \hat{\eta}_{im}\hat{\eta}_{jl} = \Delta_{ij}\Delta_{lm} \quad \mbox{and}\quad \hat{\eta}_{il}\Delta_{jm}+ \Delta_{im}\hat{\eta}_{jl} = \Delta_{ij}\hat{\eta}_{ml}.
\ee 
With this, we see that the independent tensors correspond precisely to the ones listed in \eqref{Bijred1}-\eqref{Bijred4}. Moreover, as shown in \eqref{ide3} of Appendix \ref{AI}, we can see that $B_{ijlm}^{(3)}=-B^{(1)}_{ijlm}$ and so the number of tensors reduces to three.\\
We can use the tensors  $B_{ijlm}^{(a)}$, $a= 1,2,4,$  to build more complicated objects needed for the two and three point correlators. In particular, the six indices tensors $U_{ijlmkn}$ are constructed out of combinations of  $B_{ijlm}^{(a)}$ or $T_{ijlm}^{(a)}$ which are also symmetric in the indices $(lm)$. The complete list of all the possible combinations of tensors with six indices obeying the gauge symmetry conditions is:  
\begin{align}
U^{(1)}_{ijklmn}(\vec{k})&=B^{(1)}_{ijka}(\vec{k})B^{(1)}_{lmna}(\vec{k}),   \quad U^{(9)}_{ijklmn}(\vec{k})=B^{(1)}_{ijka}(\vec{k})B^{(2)}_{lman}(\vec{k})\;, \nonumber \\
U^{(2)}_{ijklmn}(\vec{k})&=B^{(1)}_{ijna}(\vec{k})B^{(1)}_{lmka}(\vec{k}), \quad U^{(10)}_{ijklmn}(\vec{k})=B^{(1)}_{ijna}(\vec{k})B^{(2)}_{lmak}(\vec{k}), \nonumber \\
U^{(3)}_{ijklmn}(\vec{k})&=B^{(2)}_{ijan}(\vec{k})B^{(2)}_{lmak}(\vec{k}), \quad U^{(11)}_{ijklmn}(\vec{k})=B^{(1)}_{ijka}(\vec{k})B^{(2)}_{lmna}(\vec{k}),  \nonumber \\
U^{(4)}_{ijklmn}(\vec{k})&=B^{(2)}_{ijak}(\vec{k})B^{(2)}_{lman}(\vec{k}), \quad U^{(12)}_{ijklmn}(\vec{k})=B^{(1)}_{ijka}(\vec{k})B^{(2)}_{lmna}(\vec{k}),  \\
U^{(5)}_{ijklmn}(\vec{k})&=B^{(2)}_{ijan}(\vec{k})B^{(2)}_{lmka}(\vec{k}), \quad U^{(13)}_{ijklmn}(\vec{k})=B^{(2)}_{ijan}(\vec{k})B^{(1)}_{lmka}(\vec{k}), \nonumber \\
U^{(6)}_{ijklmn}(\vec{k})&=B^{(2)}_{ijak}(\vec{k})B^{(2)}_{lmna}(\vec{k}), \quad U^{(14)}_{ijklmn}(\vec{k})=B^{(2)}_{ijak}(\vec{k})B^{(1)}_{lmna}(\vec{k}), \nonumber \\
U^{(7)}_{ijklmn}(\vec{k})&=T^{(2)}_{ijlm}(\vec{k})\eta_{kna}\hat{k}_a, \qquad \; U^{(15)}_{ijklmn}(\vec{k})=B^{(1)}_{ijlm}(\vec{k})\eta_{kna}\hat{k}_a, \nonumber \\
U^{(8)}_{ijklmn}(\vec{k})&=B^{(1)}_{ijlm}(\vec{k})\Delta_{kn}, \qquad \quad U^{(16)}_{ijklmn}(\vec{k})=T^{(2)}_{ijlm}(\vec{k})\Delta_{kn}, \nonumber
\end{align}
where we have used the fact that $B^{(2)}_{ijlm}=B^{(4)}_{ijml}$. Demanding symmetry in the indices $(kn)$ and symmetrizing 
in terms of the tensors $T^{(1)}_{ijlm}$ and $T^{(2)}_{ijlm}$  defined in (\ref{Tijred1}) and (\ref{Tijred2}), we reduce the previous list to a set of independent objects:
\begin{align} \label{P14}
 P^{(1)}_{ijklmn}(\vec{k})& =T^{(1)}_{ijka}(\vec{k})T^{(1)}_{lmna}(\vec{k}),   \quad P^{(4)}_{ijklmn}(\vec{k})=T^{(1)}_{ijka}(\vec{k})T^{(2)}_{lman}(\vec{k})\;,  \\ \label{P25}
 P^{(2)}_{ijklmn}(\vec{k})& =T^{(2)}_{ijan}(\vec{k})T^{(2)}_{lmak}(\vec{k}),  \quad P^{(5)}_{ijklmn}(\vec{k})=T^{(2)}_{ijan}(\vec{k})T^{(1)}_{lmka}(\vec{k}),   \\ \label{P36}
 P^{(3)}_{ijklmn}(\vec{k})& =T^{(1)}_{ijlm}(\vec{k})\Delta_{kn},   \qquad \quad P^{(6)}_{ijklmn}(\vec{k})=T^{(2)}_{ijlm}(\vec{k})\Delta_{kn}. 
\end{align}
Those are the tensors needed to compute the $\langle \gamma \gamma \rangle$ with the contributions coming from the source vector fields. 
\section{Some useful properties}\label{AI}
Here we collect several useful properties of the projectors and the polarization vectors and tensors used in the text. The main purpose of this appendix is to demonstrate that all the tensors $T^{(n)}_{abcd}$ obey the same differential equations, and that we can extend the procedures and the results found in Ref.  \cite{Bzowski:2013sza}  to the context of parity breaking models. 
\subsection{Projectors}\label{A11}
Here we list some useful properties involving the projectors
\be
\Delta_{ab} = \delta_{ab} - \hat{k}_a\hat{k}_b \qquad \mbox{and} \qquad  \hat{\eta}_{ab} = \eta_{abc}\hat{k}_c.
\ee
These projectors obey
\ba
& k_{a}\Delta_{ab} = 0, \quad \Delta_{ac}\Delta_{cb}=\Delta_{ab}, \quad \Delta_{aa}=2,\\
& k_{a}\eta_{ab} = 0, \quad \eta_{ac}\eta_{cb}=-\Delta_{ab}, \quad \eta_{aa}=0.
\ea
The derivatives of these objects appear in the conformal Ward identities. Some useful properties involving the derivatives of $\Delta_{ab}$ are 
\begin{align}
\partial_{a}f(k) &= f'(k)\hat{k}_a,  \quad \partial_a \hat{k}_{b} =\frac{1}{k} \Delta_{ab}, \quad \partial_{a}\Delta_{ab}=-\frac{2\hat{k}_b}{k} \\
\partial_a \Delta_{bc} &= -\frac{1}{k} (\Delta_{ab} \hat{k}_{c} + \Delta_{ac} \hat{k}_{b} ), \quad k_{a} \partial_a \Delta_{bc} =0, \quad k_{b} \partial_a \Delta_{bc} = -\Delta_{ac}, \\
\nabla^2\Delta_{ab}&=-\frac{2}{k^{2}}\left(\Delta_{ab}-2\hat{k}_a\hat{k}_b\right), \quad  k_{m}\partial_{m}\partial_{l}\Delta_{ab}=-\partial_{l}\Delta_{ab}.
\end{align}
And, for the derivatives of $\hat{\eta}_{ij}=\eta_{ija}\hat{k}_a$ we have 
\begin{equation}\label{deta}
\partial_l \hat{\eta}_{ab}=\eta_{abc}\frac{\Delta_{cl}}{k}=\frac{1}{k}\left(\eta_{abl}-\hat{k}_l \hat{\eta}_{ab}\right).
\end{equation}
It is easy to see that the following identity holds
\be
\eta_{ijl}=\hat{k}_i \hat{\eta}_{jl}-\hat{k}_j \hat{\eta}_{il}+\hat{k}_l \hat{\eta}_{ij},
\ee
and if we use this identity in (\ref{deta}) we find
\be
\partial_l \hat{\eta}_{ab}=-\frac{1}{k}\left[\hat{k}_a \hat{\eta}_{lb}+\hat{k}_b \hat{\eta}_{al}\right],\quad k_{l} \partial_l \hat{\eta}_{ab} =0, \quad k_{b} \partial_l \hat{\eta}_{ab} = -\hat{\eta}_{al}, \quad \partial_{l}\hat{\eta}_{lb}=0.
\ee
which keeps the same structure of the first derivative of $\Delta_{ij}$. \\
For the second derivative of $\hat{\eta}_{ij}$ we have
\be
\nabla^2 \hat{\eta}_{ab}=-2\frac{\hat{\eta}_{ab}}{k^2}, \quad k_m\partial_m\partial_l \hat{\eta}_{ab}=-\partial_l \hat{\eta}_{ab}.
\ee
In section \ref{spectrum} we defined \eqref{Tijred1} and \eqref{Tijred2} as 
\begin{align}
T^{(1)}_{ijab}&=\frac{1}{2}\left[\Delta_{ia}\Delta_{jb}+\Delta_{ib}\Delta_{ja}-\Delta_{ij}\Delta_{ab}\right],\\
T^{(2)}_{ijab}&=\frac{i}{4}\left[\Delta_{ia}\hat{\eta}_{jb}+\Delta_{ib}\hat{\eta}_{ja}+\Delta_{ja}\hat{\eta}_{ib}+\Delta_{jb}\hat{\eta}_{ia}\right].
\end{align}
Here we didn't include the tensors associated with $B^{(3)}_{ijab}$ since using 
\begin{equation}
\hat{\eta}_{ij}\hat{\eta}_{ab}+\hat{\eta}_{ia}\hat{\eta}_{jb}=\Delta_{ia}\Delta_{jb}+\Delta_{ij}\Delta_{ab}-2\Delta_{ib}\Delta_{ja}, \label{ide3}
\end{equation}
we can show that $B^{(3)}_{ijab}=-B^{(1)}_{ijab}$.\\
Some useful properties of the tensors  $T^{(n)}_{ijab}$ are:
\be
T^{(1)}_{ljlb}=\Delta_{jb}, \quad  T^{(2)}_{ljlb}=i\hat{\eta}_{jb}.
\ee 
The following identities are useful to compute the equations coming from the SCT Ward identities 
 \begin{align}
T^{(1)}_{ajib}&=\Delta_{ai}\Delta_{jb}-T^{(1)}_{ijab},\\
T^{(1)}_{iajb}&=\Delta_{aj}\Delta_{ib}-T^{(1)}_{ijab},\\
T^{(2)}_{ajib}&=\frac{i}{2}\left(\Delta_{ia}\hat{\eta}_{jb}+\Delta_{ja}\hat{\eta}_{ib}+\Delta_{ab}\hat{\eta}_{ji}\right)-T^{(2)}_{ijab}, \label{ind2ai}\\
T^{(2)}_{iajb}&=\frac{i}{2}\left(\Delta_{ia}\hat{\eta}_{jb}+\Delta_{ja}\hat{\eta}_{ib}+\Delta_{ab}\hat{\eta}_{ij}\right)-T^{(2)}_{ijab},\label{ind2aj}.
\end{align}
To compute the relations (\ref{ind2ai}) and (\ref{ind2aj}) we have used the following identities
\begin{equation}
\Delta_{ij}\hat{\eta}_{ab}+\Delta_{bi}\hat{\eta}_{ja}+\Delta_{ab}\hat{\eta}_{ij}+\Delta_{ja}\hat{\eta}_{bi}=0.
\end{equation}
The first derivative for the tensors $T$ can be written as
\be\label{dt}
\partial_l T^{(n)}_{ijab}=-\frac{1}{k}\left[\hat{k}_i T^{(n)}_{ljab}+\hat{k}_j T^{(n)}_{ilab}+\hat{k}_a T^{(n)}_{ijlb}+\hat{k}_b T^{(n)}_{ijal}\right].
\ee
Then we obtain the following properties of the first derivative for each tensor
\begin{align}
\partial_l T^{(n)}_{ljab}&=-\frac{1}{k}\left[\hat{k}_a T^{(n)}_{ljlb}+\hat{k}_b T^{(n)}_{ljal}\right], \quad \partial_l T^{(n)}_{ijlb}=-\frac{1}{k}\left[\hat{k}_i T^{(n)}_{ljlb}+\hat{k}_j T^{(n)}_{illb}\right], \\
 \partial_l T^{(1)}_{ljab}&=\partial_j \Delta_{ab}, \quad \partial_l T^{(1)}_{ijlb}=\partial_b \Delta_{ij},\quad k_l\partial_l T^{(1)}_{ijab}=0,   \\
\partial_l T^{(2)}_{ljab}&=-\frac{i}{k}\left[\hat{k}_a\hat{\eta}_{jb}-\hat{k}_b \hat{\eta}_{aj}\right], \quad \partial_l T^{(2)}_{ijlb}=-\frac{i}{k}\left[\hat{k}_i\hat{\eta}_{jb}-\hat{k}_j \hat{\eta}_{ib}\right],\quad k_l\partial_l T^{(2)}_{ijab}=0.
\end{align}
The second derivative of  $T^{(1)}_{ijab}$ and $T^{(2)}_{ijab}$ can be written as
\begin{multline}
\partial_m \partial_l T^{(n)}_{ijab}=-\frac{\hat{k}_m}{k}\partial_l T^{(n)}_{ijab}\\
-\frac{1}{k}\left[\frac{\Delta_{im}}{k}T^{(n)}_{ljab}+\frac{\Delta_{jm}}{k}T^{(n)}_{ilab}+\frac{\Delta_{am}}{k}T^{(n)}_{ijlb}+\frac{\Delta_{bm}}{k}T^{(n)}_{ijal} \right.\\
\left.+\hat{k}_i \partial_{m}T^{(n)}_{ljab}+\hat{k}_j \partial_{m}T^{(n)}_{ilab}+\hat{k}_a \partial_{m}T^{(n)}_{ijlb}+\hat{k}_b \partial_{m}T^{(n)}_{ijal}\right].
\end{multline}
For our purposes we compute the contraction of the second derivatives 
\begin{align}
\nabla^2 T^{(1)}_{ijab}&=-\frac{4}{k^2}\left[T^{(1)}_{ijab}-\hat{k}_i \hat{k}_a \Delta_{bj}-\hat{k}_i \hat{k}_b \Delta_{aj}-\hat{k}_j \hat{k}_a \Delta_{ib}-\hat{k}_j \hat{k}_b \Delta_{ia}\right], \quad k_m\partial_m\partial_l T^{(1)}_{ijab}= -\partial_l T^{(1)}_{ijab}\\
\nabla^2 T^{(2)}_{ijab}&=-\frac{4}{k^2}\left[T^{(1)}_{ijab}-2(\hat{k}_i \hat{k}_a \hat{\eta}_{jb}+\hat{k}_i \hat{k}_b \hat{\eta}_{ja}+\hat{k}_j \hat{k}_a \hat{\eta}_{ib}+\hat{k}_j \hat{k}_b \hat{\eta}_{ia})\right], \quad k_m\partial_m\partial_l T^{(2)}_{ijab}= -\partial_l T^{(2)}_{ijab}.
\end{align}
\subsection{Polarization vectors }\label{AA1}
Here we write explicit expressions regarding the polarization vectors which are useful when contracting the various tensors involved here.  First, in general we can write a vector $\vec{k}$  in spherical coordinates as 
\be
\vec{k} = k(\sin\theta \cos \phi ,\, \sin\theta \sin \phi , \, \cos \theta).
\ee
The polarization vectors transverse to  $\vec{k}$  are obtained by solving the conditions 
\be \label{poldef}
\vec{\epsilon}_{\lambda}(\vec{k}) \cdot \vec{k} = 0, \qquad   \hat{k} \times \vec{\epsilon}_{\lambda} (\vec{k})  = - i \lambda \vec{\epsilon}_{\lambda} (\vec{k}),
\ee
The polarization vectors satisfying \eqref{poldef} obey the following properties:
\ba
&& \epsilon^{(\lambda)}_i (\hat{q}) \epsilon^{(\lambda')}_i (\hat{p}) = \frac{1}{2} \left( \frac{\vec{p}\cdot \vec{q}}{p q} -  \lambda \lambda' \right), \quad \epsilon^{(\lambda)}_i(\vec{k}) \cdot k_i = 0, \quad  \hat{\eta}_{il} \epsilon^{(\lambda)}_l  =  i \lambda  \epsilon^{(\lambda)}_i, \\
 && \epsilon^{(\lambda)}_i\epsilon^{(\lambda')}_i  =\delta_{\lambda, -\lambda'},  \quad \epsilon^{*(\lambda)}_i(\hat{k})= \epsilon^{(-\lambda)} _i(\hat{k}) = \epsilon^{(\lambda)}_i (-\hat{k}), \quad \epsilon^{(\lambda)}_i (\hat{k}) \Delta_{ij} = \epsilon^{(\lambda)}_j (\hat{k})\;,\\
&&\epsilon_{i}^{*(\lambda)}(k)\epsilon_{j}^{(\lambda)}(k)=\frac{1}{2}\left(\Delta_{ij}+i \lambda \hat{\eta}_{ij}\right)\;,  {\partial_l \epsilon^{(\lambda)}_i=-\frac{\hat{k}_i}{k}\epsilon^{(\lambda)}_l}. \label{edep}
\ea
\subsection{Polarization tensors }\label{AA2}
Polarization tensors are defined in terms of the polarization vectors as follows 
\be
\epsilon_{ij}^{(2\lambda)} (\vec{k}) = \sqrt{2} \epsilon_{i}^{( {\lambda} )} (\vec{k})  \epsilon_{j}^{( {\lambda} )} (\vec{k}). 
\ee
where $\lambda = \pm 1$. This tensor obeys the following properties:
\ba 
&&\epsilon_{ii}^{(2\lambda)} (\vec{k}) = \hat{k}_{i}\epsilon_{ij}^{(2\lambda)} (\vec{k}) =0,\, \epsilon_{ij}^{(2\lambda)} (\vec{k}) \hat{\eta}_{ij}=0\\
&&\epsilon_{ij}^{(2\lambda)*} (\vec{k}) = \epsilon_{ij}^{(-2\lambda)} (\vec{k}) = \epsilon_{ij}^{(2\lambda)} (-\vec{k}),\\  
&&\epsilon_{ij}^{(2\lambda)} (\vec{k}) \epsilon_{ij}^{(2\lambda')} (\vec{k}) = 2 \delta_{\lambda, -\lambda'}, \;  \epsilon_{ia}^{(2\lambda)} (\vec{k}) \hat{\eta}_{ja} =-i {\lambda} \epsilon_{ij}^{(2\lambda)},\\
&& {\partial_l \epsilon^{(2\lambda)}_{ij}(\vec{k})=-\frac{1}{k}\left[\hat{k}_i \epsilon^{(2\lambda)}_{lj}(\vec{k})+\hat{k}_j \epsilon^{(2\lambda)}_{il}(\vec{k})\right]\;.}
\ea
Using the previous results for the polarization tensor, we can also write a list of properties of the helicity projector tensor  
\be\Pi^{ij}_{\lambda} =  \frac{\epsilon_{-\lambda}^i(\vec{k})\epsilon_{-\lambda}^j(\vec{k})}{\sqrt{2}} \ee
that we use through the text:
\ba
\label{Piida}
&& \Pi^{ij}_{\lambda} \Delta_{il} = \Pi^{lj}_{\lambda}, \quad  \Pi^{ij}_{\lambda} \Delta_{ij} =   \Pi^{ii}_{\lambda}  = 0, \quad  \Pi^{ij}_{\lambda}  \Pi^{ij}_{-\lambda'} = \frac{1}{2} \delta_{\lambda, \lambda'}\\ \label{Piidb} 
&&  \Pi^{ia}_{\lambda}\hat{\eta}_{ja} =i\lambda  \Pi^{ij}_{\lambda}, \quad  \Pi^{ij}_{\lambda}\hat{\eta}_{ij} =0, \\
&& \Pi^{ij}_{\lambda}B^{(1)}_{ijlm}= 2 \Pi^{lm}_{\lambda} , \quad \Pi^{ij}_{\lambda}B^{(2)}_{ijlm}=- 2 i \lambda \Pi^{lm}_{\lambda}, \label{pb1} \\
&& \Pi^{ij}_{\lambda}B^{(3)}_{ijlm}=- 2 \lambda^2 \Pi^{lm}_{\lambda}, \quad \Pi^{ij}_{\lambda}B^{(4)}_{ijlm}=- 2 i \lambda \Pi^{lm}_{\lambda},\label{pb2}\\
&& \Pi^{ij}_{\lambda}T^{(1)}_{ijlm}=  \Pi^{lm}_{\lambda}, \quad \Pi^{ij}_{\lambda}T^{(2)}_{ijlm}=  \lambda \Pi^{lm}_{\lambda}, \label{pt123}\\
&&2\hat{E}^l\hat{E}^m\Pi^{la}_{\lambda}\Pi^{am}_{-\lambda'}=\delta_{\lambda,\lambda'}\hat{E}^l\hat{E}^m\Delta_{lm}\label{eepipi}\;. \label{p3}
\ea
 
\section{Conformal Ward identities and triple K integrals}\label{fm}
In general the conformal Ward identities can be written as 
\ba
\left[-3(N-1) + \sum_{a=1}^{N} \left( \Delta_{a}  -  \vec{k}_a \cdot   \frac{\partial}{\partial \vec{k}_a }  \right)  \right] \langle \sigma(\vec{k}_1)\cdots  \sigma(\vec{k}_s) v_{i} (\vec{k}_{s+1})\cdots v_{j} (\vec{k}_N)\rangle' = 0,
\ea
for the Dilatation ward identity. And for the Conformal Ward identity we have

\begin{multline}
\left[ \sum_{a=1}^{N} 2(\Delta_{a} -3) \partial_{k_a^i} + D^{i}_a  \right] \langle \sigma(\vec{k}_1)\cdots  \sigma(\vec{k}_s) v_{i_1} (\vec{k}_{s+1})\cdots v_{i_N} (\vec{k}_N)\rangle'   \\      -2 \sum_{p=s+1}^{N}  \Sigma^{i j_p}{}_{i_p} \langle \sigma(\vec{k}_1)\cdots  \sigma(\vec{k}_s) v_{i_1} (\vec{k}_{s+1}) \cdots v_{j_p} (\vec{k}_{p}) \cdots v_{i_N} (\vec{k}_N)\rangle'  =0,\nonumber
\end{multline}
where
\ba
 D^{i}_a =  k^{i}_{a} \partial^2_{k_a} - 2(\vec{k}_a\cdot \vec{\partial}_{k_a}) \partial_{k^{i}_a} \quad \mbox{and} \quad
 \Sigma^{i j_p}{}_{i_p} = \delta^{i j_p}\partial_{k^{i_p}_p} -   \delta^{i}_{i_p}\partial_{k^{j_p}_p}. 
\ea
for more details on the conformal ward identities you can check \cite{Almeida:2017lrq, Maldacena:2011nz}. In order to illustrate how the method used in \cite{Bzowski:2013sza} works, we compute the equations for a 3 point correlation function consisting in rank-2 tensor and two scalars. We will assume the tensor to be traceless, divergenceless and symmetric since the tensor perturbations also satisfy these conditions. To start we propose a form for the correlation function to be
\begin{equation}
\langle t_{ij}(k_1)O_1(k_2)O_2(k_3)\rangle'=T_{ijab}(k_1)k_2^ak_2^b A(k_1,k_2,k_3).
\end{equation}
And here we will keep the tensor $T_{ijab}$ generic since the two objects $T^{(n)}_{ijab}$, satisfy the same equations for they derivatives as shown in (\ref{dt}) and they all have the same properties of trace free and symmetry in the first pair and the second pair of indices and divergence zero for all the indices. With this, the SCT Ward identity can be written like
\begin{multline}
\sum_{a=1}^{N} \left[2(\Delta_{a} -3) \partial_{k_a^l} +k^{l}_{a} \partial_{k_a^m}\partial_{k_a^m} - 2k_a^m \partial_{k_a^m} \partial_{k^{l}_a}\right] T_{ijab}(k_1)k_2^ak_2^b A(k_1,k_2,k_3) \\
-2k_2^ak_2^b\left[\partial_{k_1^i}(T_{ljab}(k_1)A(k_1,k_2,k_3))-\delta_{li}\partial_{k_1^m}(T_{mjab}(k_1)A(k_1,k_2,k_3))\right] \\
-2k_2^ak_2^b\left[\partial_{k_1^j}(T_{ilab}(k_1)A(k_1,k_2,k_3))-\delta_{lj}\partial_{k_1^m}(T_{imab}(k_1)A(k_1,k_2,k_3))\right]=0.
\end{multline} 
After an extensive calculation and using the properties described in \ref{AI} we find the following result
\begin{equation}\label{tssf}
T_{ijab}k_2^ak_2^bk_1^l\mathcal{K}_{13}A+T_{ijab}k_2^ak_2^bk_2^l\mathcal{K}_{23}A+4T_{ijlb}k_2^b\mathcal{L}A-2k_2^ak_2^b\left[\hat{k}_1^iT_{ljab}+\hat{k}_1^jT_{ilab}\right]\frac{A}{k_1}(\Delta_1-3)=0, \\
\end{equation}
where $\mathcal{K}_{ij}$ are the primary ward identities used in \cite{Bzowski:2013sza} defined, for our particular case, as
\begin{equation}\label{psctwi}
\mathcal{K}_{ij}=2(\Delta_i-2)\frac{1}{k_i}\partial_i-\partial^2_i-2(\Delta_j-2)\frac{1}{k_j}\partial_j+\partial^2_j.
\end{equation} 
By the other hand, $\mathcal{L}$ is one of the secondary SCT Ward identity written like
\begin{equation}
\mathcal{L}=(\Delta_2-4)-k_2 \partial_2-(\vec{k}_2 \cdot \vec{k}_1)\frac{1}{k_1^2}(\Delta_1-k_1\partial_1).
\end{equation}
In order to satisfy \ref{tssf} the following equations must hold
\begin{align}
\mathcal{K}_{13}A&=0, \quad \mathcal{K}_{23}A=0, \\
 \mathcal{L}A&=0, \quad \Delta_1=3. 
\end{align}
Notice that for this case the primary equations (which are the equations of interest in this work) are homogeneous, but in more complicated cases they could not be. For instance, doing a similar procedure for $\langle \gamma_{ij}^{(1)}\zeta^{(1)}\zeta^{(0)}\rangle$ we arrive at the following set of primary SCT Ward identities
\begin{align}
\mathcal{K}_{12}A_1&=0, \quad \quad \mathcal{K}_{13}A_1=0,  \\
\mathcal{K}_{12}A_2&=0, \quad \quad \mathcal{K}_{13}A_2=0, \\
\mathcal{K}_{12}A_3&=4 A_1, \quad  \mathcal{K}_{13}A_3=0, \\
\mathcal{K}_{12}A_4&=0, \quad \quad \mathcal{K}_{13}A_4=0,  \\
\mathcal{K}_{12}A_5&=2A_4, \quad  \mathcal{K}_{13}A_5=0,
\end{align}
which are solved by the triple-K integrals described in section \ref{bispectrum}. For more details about the primary and secondary conformal Ward identities, the reader is referred to \cite{Bzowski:2013sza}.

\bibliographystyle{unsrt} 
\bibliography{bibli} 

\end{document}

%% file: shortcuts.tex
\newcommand{\ba}{\begin{eqnarray}}
\newcommand{\ea}{\end{eqnarray}}
\newcommand{\be}{\begin{equation}}
\newcommand{\ee}{\end{equation}}
\newcommand{\bea}{\begin{eqnarray}}
\newcommand{\eea}{\end{eqnarray}}
\newcommand{\beq}{\begin{equation}}
\newcommand{\eeq}{\end{equation}}
\newcommand{\beqar}{\begin{eqnarray}}
\newcommand{\eeqar}{\end{eqnarray}}
\newcommand{\beqars}{\begin{eqnarray*}}
\newcommand{\eeqars}{\end{eqnarray*}}
\newcommand{\bc}{\begin{center}}
\newcommand{\ec}{\end{center}}
\newcommand{\ben}{\begin{enumerate}}
\newcommand{\een}{\end{enumerate}}
\newcommand{\bit}{\begin{itemize}}
\newcommand{\eit}{\end{itemize}}
\newcommand{\bw}{\begin{widetext}}
\newcommand{\ew}{\end{widetext}}
\newcommand{\bcl}{\begin{columns}}
\newcommand{\ecl}{\end{columns}}